\RequirePackage{fix-cm}
\documentclass[twocolumn,epjc3]{svjour3}  
\smartqed  % flush right qed marks, e.g. at end of proof
\RequirePackage{graphicx}
\journalname{Eur. Phys. J. C}
\usepackage{cancel}
\usepackage{cite} %A: para contraer referencias
\usepackage[nottoc]{tocbibind}
\usepackage{amsmath}%A
\usepackage{hyperref}
\usepackage{cite}%A
\usepackage{amsmath,amssymb}%A
\usepackage{color}%A
\usepackage{subfigure}
\usepackage{float}
\usepackage{multicol}
\usepackage{hyperref,amssymb,amsmath,mathrsfs,bm,graphicx}
\usepackage{multirow}
\usepackage{array}
\usepackage{float}
\usepackage{dblfloatfix}
\usepackage{placeins}
\usepackage[dvipsnames]{xcolor}

\begin{document}
\sloppy
\title{Quasi Normal Modes of hairy black holes at higher--order WKB
approach.}
%\subtitle{Do you have a subtitle?\\ If so, write it here}

%\titlerunning{Short form of title}        % if too long for running head

\author{R. Avalos \thanksref{e2,addr3}
        \and E. Contreras\thanksref{e3,addr3}
        }
\thankstext{e2}{e-mail: 
\href{mailto:ravalos@usfq.edu.ec}{\nolinkurl{ravalos@usfq.edu.ec}}}
\thankstext{e3}{e-mail: 
\href{mailto:econtreras@usfq.edu.ec}{\nolinkurl{econtreras@usfq.edu.ec}}}
%\thankstext{e4}{e-mail: 
%\href{mailto:zdenek.stuchlik@physics.slu.cz}{\nolinkurl{zdenek.stuchlik@physics.slu.cz}}}

\institute{
Departamento de F\'isica, Colegio de Ciencias e Ingenier\'ia, Universidad San Francisco de Quito,  Quito, Ecuador.\label{addr3}
%\and Research Centre for Theoretical Physics and Astrophysics,
%Institute of Physics, Silesian University in Opava, CZ 746 01 Opava, Czech Republic.\label{addr4}
}

\date{Received: date / Accepted: date}
% The correct dates will be entered by the editor

\maketitle

\begin{abstract}
In this work, we implement the $13^{th}$ order semi-analytical WKB method to explore the stability of hairy black holes obtained in the framework of Gravitational Decoupling. In particular, we perform a detailed analysis of the frequencies of the quasi-normal modes as a function of the primary hair of the solutions with the aim to bound their values. We explore a broad interval in a step of 0.1 of the hair parameters. 
We find that except for some cases where the method is expected to have poor accuracy, all the solutions seem to be stable and the role played by the primary hair is twofold: to modulate the damping factor of the perturbation and to decrease the frequency of its oscillation. 
\end{abstract}
%%%%%%%%%%%%%%%%%%%%%%%%%%%%%%%%%%%%%%%%%%

\section{Introduction}
Although non--hair conjecture states that black holes (BH) are the simplest objects in nature described only by a few parameters, namely its mass, charge, and angular momentum, a real BH is far from being isolated. Indeed, black holes are surrounded by galactic nuclei, stars, planets, etc, so they are always in a perturbed state \cite{Konoplya:2011qq}. In this regard, to analyze the stability of BH's we have to start with the study of their perturbation.

For the study of how a BH responds to perturbation, we can consider either perturbing the BH background (the spacetime metric) or adding extra fields to the BH spacetime. In any case, the equation describing the perturbation of the BH reduces to a radial like--Shr\"odinger equation in which complex frequencies solutions correspond to the quasinormal modes (QNM) of the BH. More precisely, the real part of the QNM frequencies, $Re(\omega)$, corresponds to the frequency of oscillations, and the imaginary part, $Im(\omega)$, is related to the damping factor associated with the loss of energy through gravitational radiation. It is worth mentioning that if $Im(\omega)>0$ the perturbation grows exponentially leading to instabilities in the system so that a stable solution will be that which $Im(\omega)<0$. 

The computation of the QNM modes can be performed through a variety of methods (for an incomplete list see \cite{Konoplya:2022tvv,Churilova:2021nnc,Konoplya:2020jgt,Konoplya:2020bxa,Konoplya:2019nzp,Rincon:2021gwd,Panotopoulos:2020mii,Rincon:2020cos,Rincon:2020iwy,Rincon:2020pne,Xiong:2021cth,Zhang:2021bdr,Panotopoulos:2019gtn,Lee:2020iau,Churilova:2019qph,Oliveira:2018oha,Panotopoulos:2017hns} and references therein, for example). However, in this work, we shall use the recently developed  WKB approximation to the $13^{th}$ order which has brought the attention of the community \cite{Konoplya:2019hlu}. It should be emphasized that as the computation of QNM modes can only be performed semi--analytically, the set of any free parameter appearing in the BH solution is compulsory. As a consequence, the computation of QNM modes allows the definition of the parameter space of any solution by demanding $Im(\omega)<0$. This strategy has been applied in Ref. \cite{Konoplya:2018ala} with the aim to bound the free parameters in the construction of stable traversable wormholes.

In this work, we explore the stability through the computation of the QNM by the $13^{th}$ order WKB method of hairy black holes obtained by Gravitational decoupling in Ref. \cite{Ovalle:2020kpd}. The main goal is to bound in the values taken by the ``primary'' hairs of the solution. It is worth mentioning that, during the writing of this work, there have been reported two independent studies on the QNM modes for the same models \cite{Cavalcanti:2022cga,Yang:2022ifo}. However, in our analysis, we make an extensive study on the frequencies as a function of the primary hairs of the solution that has not been reported before.

This work is organized as follows. In the next section, we review the main aspects related computation of the QNM associated to perturbation of a BH. In section \ref{models}, we introduce the hairy BH models in which we base our analysis. Section \ref{results} is devoted to the analysis of the results obtained and in the last section we conclude the work.

\section{QNM by the WKB approximation}\label{QNMi}
Let us consider a static and spherically symmetry line element satisfying the Schwarzschild condition, namely
\begin{eqnarray} \label{metric}
ds^2=f(r)dt^2 -\frac{dr^2}{f(r)}-r^2 (d\theta^2+\sin^2{\theta} d\phi^2),
\end{eqnarray}
with  $f(r)$ the so--called lapse function encoding the information of the BH spacetime. Next, the perturbation of the BH can be performed by adding test fields (Klein-Gordon or Dirac fields) to the background or by perturbing the spacetime itself. In any case, the perturbation equation can be reduced to a like--Schr\"{o}dinger equation of the form
\begin{eqnarray} \label{schrodinger}
\bigg( \frac{d^2}{dr_*^2} +\omega ^2 -V(r_*) \bigg)\chi(r_*)=0,
\end{eqnarray}
where $r_*$ is the tortoise radial coordinates
\begin{eqnarray}
\frac{dr_*}{dr}=\frac{1}{f(r)},
\end{eqnarray}
and $V(r)$ is an effective potential, which for axial perturbations takes the form
\begin{eqnarray} \label{veff}
V_L (r) = f(r) \bigg( \frac{L (L+1)}{r^2} +f'(r)\frac{(1-s^2)}{r} \bigg),
\end{eqnarray}
where $s=0,1,2$  is the spin of the perturbing field. In this work, we shall impose $s=0$ (scalar field). Besides, $\omega$ represent the frequency of the QNM and has a real and an imaginary part, namely $\omega=Re(\omega)+iIm(\omega)$. 

Several strategies can be implemented to obtain the QNM  frequencies. However, in this work we shall implement the WKB approximation first introduced in Ref. \cite{Schutz:1985km} to study scattering around BH's, given its similarity with the one--dimensional Schr\"{o}dinger equation with a potential barrier. Now, as the problem demands that both the ``reflected'' and ``transmitted'' waves of the scattering problem have comparable amplitudes, the problem reduces to implement the WKB method to high orders around the top of the potential. It can be shown that the $13^{th}$ WKB order formula reads
\begin{eqnarray} \label{WKB}
i \frac{\omega ^2 -V_0}{\sqrt{-2V_0''}}-\sum_{j=2}^{13}\Lambda_j=n+\frac{1}{2},
\end{eqnarray}
where $V_0$ is the maximum height of the potential and $V_0''$ is its second derivative with respect to the tortoise coordinate evaluated at the radius where $V_{0}$ reaches a maximum. The values $\Lambda_j$ are corrections that depend on the value of the potential and higher derivatives of it at the maximum. The exact expressions for the terms $\Lambda_j$ are too long to be shown here but can be found in  \cite{Konoplya:2019hlu}.

\section{Hairy black holes}\label{models}
In this section we briefly describe the hairy black holes obtained in Ref. \cite{Ovalle:2020kpd} by the Gravitational Decoupling (GD) through 
the Minimal Geometric Deformation (MGD) extended (for details about GD and MGD see \cite{Heras:2022aho,Maurya:2022hav,Azmat:2022bbc,Maurya:2022uqu,Andrade,Ovalle:2022yjl,Ovalle:2022eqb,Maurya:2021yhc,Zubair:2021lgt,daRocha:2021sqd,Azmat:2021kmv,Maurya:2021fuy,daRocha:2020rda,Meert:2021khi,Sultana:2021cvq,Tello-Ortiz:2021kxg,Maurya:2021aio,daRocha:2020jdj,Tello-Ortiz:2020euy,Ovalle:2020fuo,daRocha:2020gee,Meert:2020sqv,Tello-Ortiz:2020ydf,Maurya:2020rny,Maurya:2020gjw,Ovalle:2019qyi,daRocha:2019pla,Heras:2019ibr,Maurya:2019wsk,Gabbanelli:2019txr,Estrada:2019aeh,Ovalle:2019lbs,Maurya:2019hds,Hensh:2019rtb,Fernandes-Silva:2019fez,Leon:2019abq,Singh:2019ktp,Maurya:2019noq,Gabbanelli:2018bhs,Ovalle:2018umz,Fernandes-Silva:2018abr,Morales:2018nmq,Estrada:2018zbh,Heras:2018cpz,Panotopoulos:2018law,Morales:2018urp,Estrada:2018vrl,Ovalle:2018ans,Ovalle:2017wqi,Ovalle:2017fgl}). As it is shown in
\cite{Ovalle:2020kpd}, the solutions satisfy either the strong (SEC) and the dominant (DEC) energy conditions outside the horizon which make them attractive geometries in the modeling of suitable BH-hair systems.

\subsection{Model 1}
This black hole is characterized by a metric function of the form
\begin{eqnarray} \label{M1}
f(r)=1-\frac{2M}{r}+\alpha \bigg( e^{-r/M} -\frac{2M}{r} e^{-2} \bigg),
\end{eqnarray}
where $\alpha$ is the decoupling parameter which connects \eqref{M1} with the Schwarzschield black hole which is obtain in the limit $\alpha\to0$. Note that the event horizon is located in $r_H=2M$ which equals the Schwarzschild horizon. As can be seen in Ref. \cite{Ovalle:2020kpd}, this solution satisfies the SEC outside the horizon.

\subsection{Model 2}
In this case, the metric function takes the form
\begin{eqnarray} \label{M2}
f(r)&=&1-\frac{2M}{r}+\bigg(1+\frac{\alpha}{2e^2} \bigg) \nonumber\\
&& +\frac{4 \alpha M^2}{e^2 r^2}-\frac{\alpha M}{r}e^{-r/M},
\end{eqnarray}
As in the previous case, this solution reduces to the Schwarzschild BH when $\alpha=0$. Besides, the horizon radius is located at $r_{H}=2M$. As shown in Ref. \cite{Ovalle:2020kpd}, this solution satifies the DEC outside the horizon.

\subsection{Model 3}
This black hole is characterized by a metric function of the form
\begin{eqnarray} \label{M3}
f(r)&=&1-\frac{2M+\ell_0}{r}+ \frac{2\ell_0 M}{r^2} \nonumber\\
&&-\frac{\alpha M e^{-r/M}}{r^2} \big( r-\ell_0 e^{\frac{r-\ell_0}{M}} \big),
\end{eqnarray}
and has the event horizon is  in $r_H=\alpha \ell=\ell_0$, being $\ell$ a parameter that relates $\alpha$ and $\ell_0$. For the rest of the work we will fix $\ell=1$ which means $\alpha=\ell_0$. This solution satisfies the DEC.

\subsection{Model 4}
This black hole is characterized by a metric function of the form
\begin{eqnarray} \label{M4}
f(r)&=&1-\frac{2M+\ell_0}{r} \nonumber\\
&&-\frac{\alpha M e^{-r/M}}{r^2} \bigg( r-(2M+\ell_0) e^{\frac{r-(2m+\ell_0)}{M}} \bigg),
\end{eqnarray}
and has the event horizon is  in $r_H=\alpha \ell=\ell_0$, being $\ell$ a parameter that relates $\alpha$ and $\ell_0$. For the rest of the work we will fix $\ell=1$ which means $\alpha=\ell_0$. This solution satisfies the DEC.

Before concluding this section, a couple of comments are in order. First, it is worth mentioning that in \cite{Ramos:2021jta} the geodesic analysis of the set of metric describing models 1-4 has been performed. In such a work, the authors not only studied the basic stuff related to the motion of  massive and massless particles around the hairy black hole but they explored their potential as a mimickers of rotating black holes based on available data. This was achieved by comparing the radius of the innermost circular orbits (ISCO) of the static solution with that obtained from the Kerr solution. The relation between both ISCO radius was used to bound the values of the hairs as a function of the spin parameter of the rotating solution. The results obtained allow to consider the black holes described by the metrics here as mimickers of the systems ARK564 and NGC1365. The spin parameter of both systems was derived form relativistic reflection fitting of SMBH X-ray as reported in
\cite{Risaliti:2009dn,Risaliti:2013cga,Brenneman:2012ws}. Second, the set of metric considered here was used as the seed for the construction of a rotating black hole by following the Gravitational Decoupling approach for stationary space--times \cite{Contreras:2021yxe}. Although in that work the shadow cast by this solution was found, an extensive analysis can be perform in order to compare the results with the EHT data. Indeed, we can use the strategy followed in \cite{Jusufi:2020odz} where the authors constraint Einstein-Yang-Mills parameter via frequency analysis of the quasi periodic normal oscillations and the EHT data of shadow cast by the M87 super massive BH. In the same direction, the results obtained in \cite{Contreras:2021yxe}  based on the set of metric we are assuming in the present work, could be used to constraint the value of the hair associated with each of them. However, although this treatment is out of the scope of this work, we can estimate the typical size of the shadow, $\mathcal{R}_{s}$, of the M87 supermassive BH based on the set of the static metrics here as an approximation. To this end, we proceed as follows. First, the angular diameter of the BH shadow, $\theta_{s}$, can be expressed as \cite{Jusufi:2020odz}
\begin{eqnarray}
\theta_{s}=2\times9.87098\times10^{-6}\mathcal{R}_{s}
\left(\frac{M}{M_{\odot}}\right)\left(\frac{1kpc}{D}\right)\mu as,
\end{eqnarray}
where $M$ and $D$ are the mass and the distance of the BH, respectively. For the M87 BH, $\theta_{s}=(42\pm3)\mu as$, $M=6.5\times10^{9}M_{\odot}$ and $D=16.8Mpc$ so that $\mathcal{R}_{s}=5.49874$.
Second, from \cite{Jusufi:2019ltj,Jusufi:2020dhz} it is known that the relationship between the shadow and the radius of the photon sphere $r_{0}$ is given by
\begin{eqnarray}\label{RSha}
\mathcal{R}_{s}=\frac{r_{0}}{f(r_{0})}
\end{eqnarray}
with $f$ the lapse function under consideration. The radius of the photon sphere of the metrics under consideration was obtained numerically and are shown in figure 7 of \cite{Ramos:2021jta} for $\{\alpha,\ell_{0}\}\in(0,3)$. Using this data in (\ref{RSha}), we obtain $\mathcal{R}_{s}$ as function of the primary hairs as is shown in Fig. \ref{RSha}. Note that, the model 2 is the only one that cannot be used to mimic the shadow of the M87 supermassive BH. It should be emphasized that comparison we are doing here must be taken as an approximation.

\begin{figure*}[hbt!]
\centering
\includegraphics[width=0.4\textwidth]{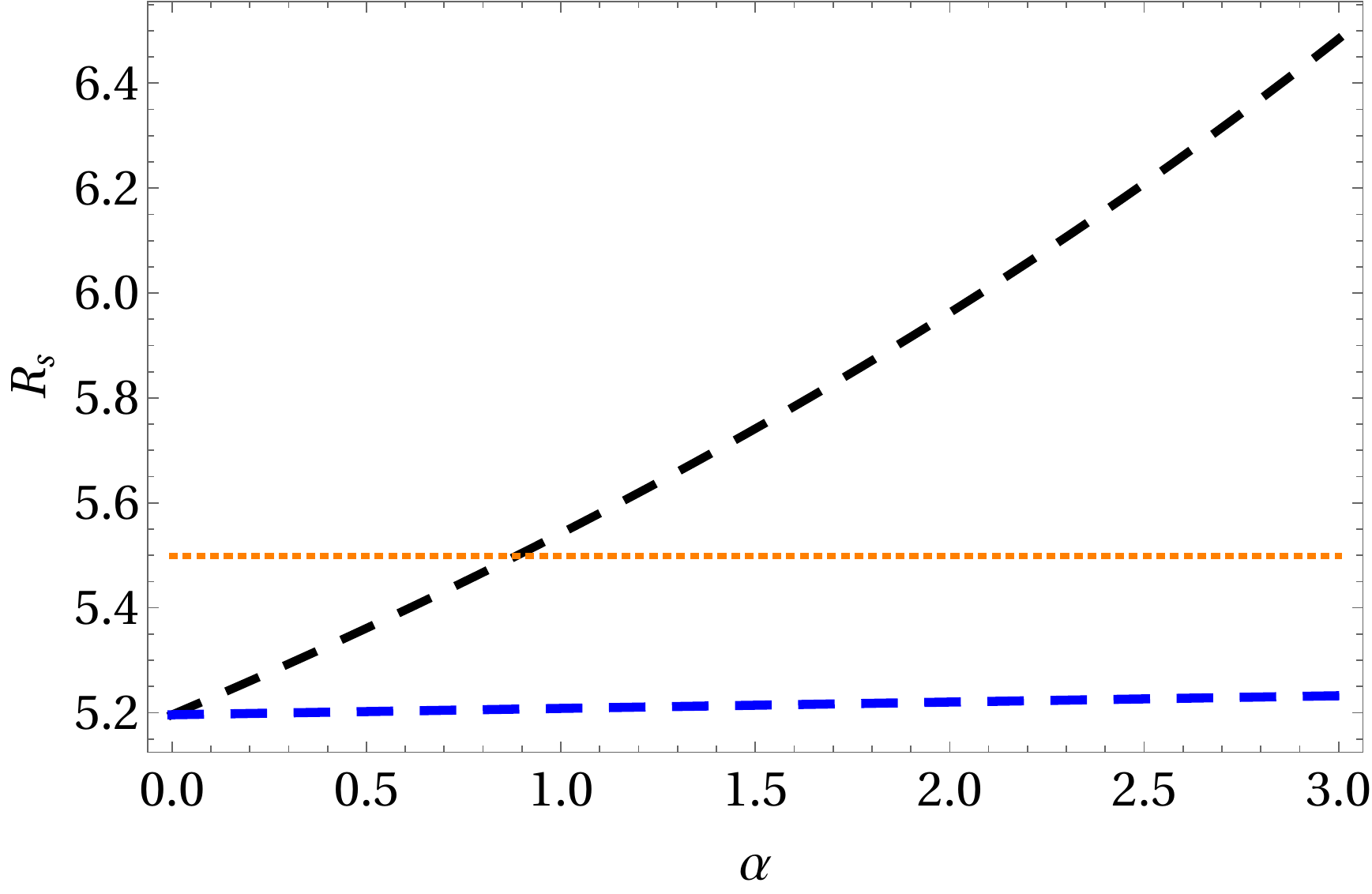}  \
\includegraphics[width=0.4\textwidth]{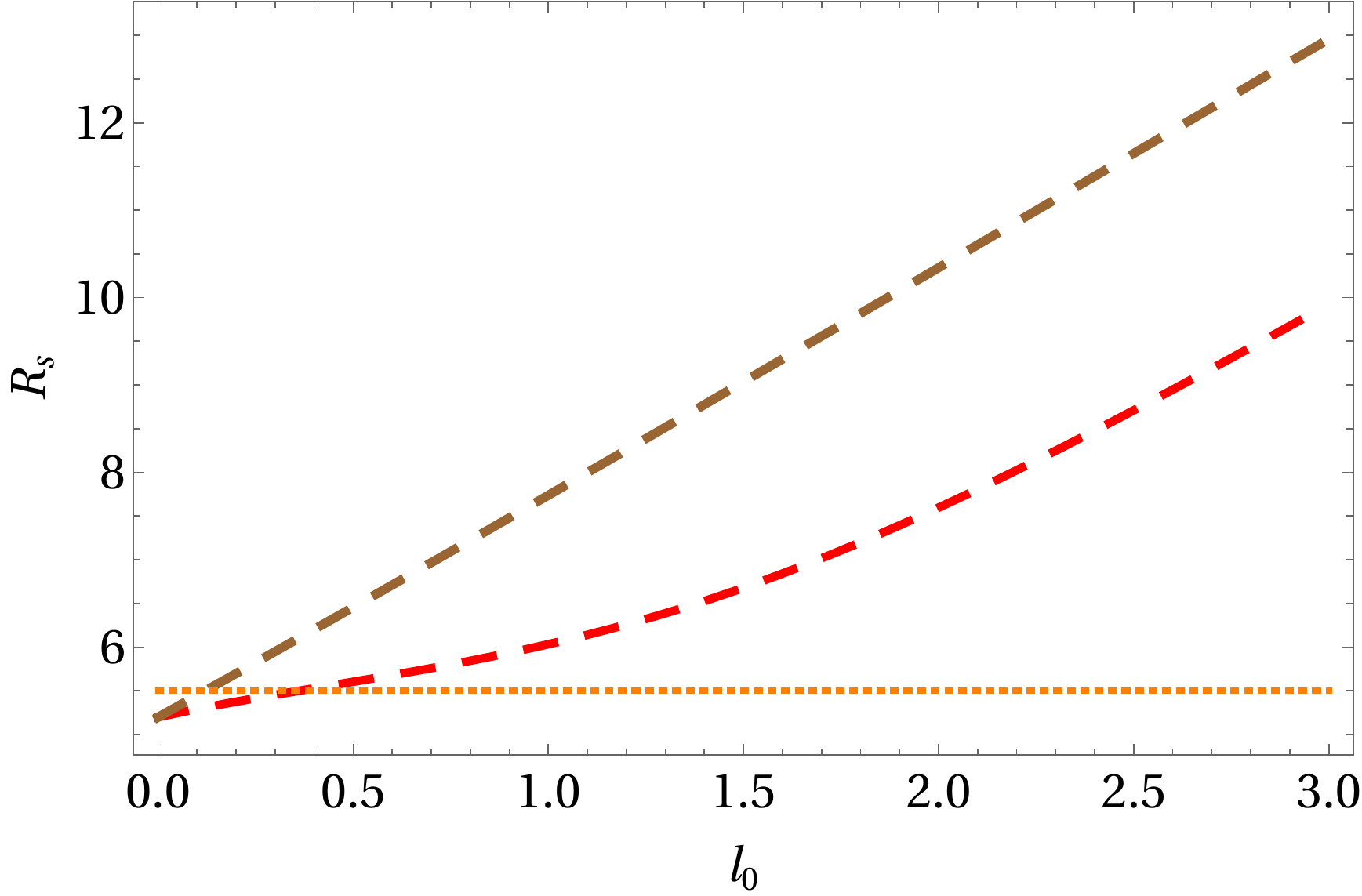}  \
\caption{\label{RSha}
Typical radius of the shadow, $\mathcal{R}_{s}$ as function of $\alpha$ for models 1 (black line) and 2 (blue line) in the left panel and as a function of $\ell_{0}$ for model 3 (red line) and 4 (brown line) in the right panel. The horizontal orange line corresponds to $\mathcal{R}_{s}=5.49874$ for M87 BH as discussed in the text.}
\end{figure*}

It is worth mentioning that the primary hairs, namely $\alpha$ in Models 1 and 2 and $\ell_{0}$ in Models 3 and 4, can take arbitrary values in principle. For example, in Refs. \cite{Ovalle:2020kpd,Ramos:2021jta}  we have taken $\{\alpha,\ell_{0}\}\in(0,1)$. In this work we explore the stability of the solution in a larger interval. Although we have studied the behaviour for $\alpha,\ell_{0}\in(-50,50)$, here we show the results in the interval $(-10,10)$ given that, essentially, it  contains all the information we require.

\section{Results and Discussion}\label{results}
In this section, we shall discuss the results obtained by the implementation of the WKB method in the models described in the previous sections. In all of the cases have plotted $Im(\omega)$ as a function of the primary hair of the BH with the aim to explore if there is a change in the imaginary part of the frequency for some of their values. All the results are shown in Figs. (\ref{M1}), (\ref{M2}), (\ref{M3}) and (\ref{M4}) for Models 1, 2, 3 and 4 respectively.  All the computations have been performed setting the multipole number $L$ and varying the overtone $n=0,1,2,3,4$. For each model we have a plot for each $L=2,3,4,5,6,7,8,9,10$. It is worth mentioning that in all the plots we have taken a step of $0.1$ for the hair parameters, namely $\alpha,\ell_{0}$. Higher precision is possible but the computational time increases considerably.

In Fig. \ref{M1} for $L=2$, we note an increasing in the value of $Im(\omega)$ and then an oscillatory behaviour around $\alpha=5$. Even more, there is a change of sign for $n=3$ and $n=4$. The same oscillatory behaviour is observed for $L=3,4,5,6$ and $n=2,3,4$. However, for $L=7,8,9,10$ the function increases monotonously up to a certain $\alpha\approx5$ where the frequency reaches a maximum and then decreases and converges to a certain constant value. It is worth mentioning that it is claimed that the method has high accuracy for small $n$ and large $L$ so that, the change of sign for $n=3,4$ for $L=2$ could be associated with the lack of precision for the values under study and not to instabilities of the BH. Similarly, we could conclude that the oscillatory behavior is a result of the low precision of the method for large $n$ and small $L$. Finally, the damping of the signal given by $e^{-iIm(\omega)}$, decreases as $\alpha$ grows (except in the oscillatory interval). Accordingly, as $\alpha$ increases, the oscillation dominates at late times. Furthermore, the frequency of the the oscillations (the $Re(\omega)$) decreases when $\alpha$ increases as shown in Tables \ref{table:num}, \ref{table:num2}, and \ref{table:num3}. Also, we note that
as $\alpha$ takes bigger values, the $Re(\omega)$ converges to the same values for the different overtones $n$ and a fixed $L$. In summary, as the parameter associated with the primary hair increases the signal becomes less damped and the dominant oscillatory behavior at late times becomes ``monochromatic'' for each multipole number $L$.

In Fig \ref{M2} we show the results for Model 2. Note that for $L=2$ the frequency remains constant for $n=0,1,2$. Besides, the value of the frequency decreases as $n$ increases. For $n=3,4$ the frequency is a decreasing function and undergoes an oscillatory behavior around $\alpha\approx10$. In contrast, for $L=3,4,5,6,7,8$, $Im(\omega)$ is constant for $n=0,1,2$ but a maximum around $\alpha\approx0$ for $n=3,4$. For $L=9,10$ the frequency is almost constant for every value of the overtone under consideration. However, in contrast to what occurs in Model 1, the values of $Re(\omega)$ are different for a fixed multiple number $L$ and different overtones $n$. Even more, their separation increase as the primary hair grows as shown in tables \ref{table:num}, \ref{table:num2}, and \ref{table:num3}. In any case, the results indicate that Model 2 is stable for all the values of the hair considered here. Based on these results, we conclude that for this model, the damping of the signal is almost the same for each value of the primary hair but the frequency of oscillations at late times depends critically on the value of $\alpha$.

The results for Model 3 are shown in Fig. \ref{M3}. We note that, except for $L=2,3$ where there is an oscillatory behavior for $n=3,4$, the frequency increases monotonously. Besides, $Im(\omega)$ is always negative indicating that this model is stable for each value of the primary hair under consideration. Again, as in Model 1, both the damping of the signal and $Re(\omega)$ decreases  as $\alpha$ grows (except in the oscillatory interval)

In Fig. \ref{M4} we show the frequency for Model 4. The behavior is similar to that seen in Models 1 and 3. The only difference is that, in this case, there is not any oscillatory behavior for any values of $n$ or $L$. 

Another point that deserves discussion is the values taken by $Im(\omega)$ for a fixed multipole, $L$, and different values of the overtone $n$. A well--established point is that for a fixed $L$, the absolute value of the imaginary part of the QNM frequency increases as $n$ grows \cite{Konoplya:2011qq,Konoplya:2018ala}. This behavior is shown by Models 2, 3, and 4 for all the values of the primary hairs under consideration. In this regard, as $n$ increases, the damping is strong. In Model 1, the oscillatory behavior for $L=2,3$ and $n=3,4$ violates this tendency for a certain value of $\alpha$ but as we stated before this should be related to the lack of accuracy of the method.

In summary, we conclude that the role played by the
primary hair in the BH is twofold: to modulate the damping factor of the perturbation and to decrease the frequency of the dominant oscillations at late times.

Before concluding this section we would like to emphasize that the results obtained here can be matched with observational data thorough the relationship between the real part of the QNM and the the typical radius of the shadow $\mathcal{R}_{s}$. Indeed, in Ref. \cite{Cuadros-Melgar:2020kqn} the authors found a relationship between the QNM frequencies and the metric evaluated at the radius of the photon sphere at third order WKB. It should be interesting to explore this relationship to higher orders with the aim to apply this results with the finding here. However, this issue is out the scope of this paper and could be explore in a future work.

\begin{table*}[htb!]
\centering
\begin{tabular}{|m{2em}|m{2em}|m{4em}|m{4em}|m{4em}|m{4em}|m{4em}|m{4em}|m{4em}|m{4em}|} 
 \hline
 \multicolumn{2}{|c|}{ } & \multicolumn{2}{|c|}{Model 1}  & \multicolumn{2}{|c|}{Model 2} & \multicolumn{2}{|c|}{Model 3} & \multicolumn{2}{|c|}{Model 4}\\  
 \hline
 L & $n$ & \multicolumn{1}{|c|}{Re $\omega$}  & \multicolumn{1}{|c|}{Im $\omega$} & \multicolumn{1}{|c|}{Re $\omega$}  & \multicolumn{1}{|c|}{Im $\omega$} & \multicolumn{1}{|c|}{Re $\omega$}  & \multicolumn{1}{|c|}{Im $\omega$}& \multicolumn{1}{|c|}{Re $\omega$}  & \multicolumn{1}{|c|}{Im $\omega$}\\
\hline
6 & 0 & 1.17395 & -0.08208 & 1.24877 & -0.09627 & 1.07828 & -0.06557 & 0.82851 & -0.06307\\ 
& 1 & 1.16863 & -0.24669 & 1.24019 & -0.28961 & 1.07411 & -0.19699 & 0.82337 & -0.18975\\ 
& 2 & 1.15831 & -0.41255 & 1.61297 & -0.48535 & 1.06586 & -0.32919 & 0.81332 & -0.31804\\
\hline
7 & 0 & 1.35425 & -0.08206 & 1.44067 & -0.09625 & 1.24398 & -0.06557 & 0.95578 & -0.06306\\ 
& 1 & 1.34961 & -0.24654 & 1.43322 & -0.28937 & 1.24036 & -0.19691 & 0.95131 & -0.18958\\ 
& 2 & 1.34054 & -0.41198 & 1.61297 & -0.48429 & 1.23317 & -0.32885 & 0.94253 & -0.31731\\
\hline
8 & 0 & 1.53458 & -0.08206 & 1.63259 & -0.09624 & 1.40970 & -0.06557 & 1.08307 & -0.06305\\ 
& 1 & 1.53047 & -0.24644 & 1.62601 & -0.28920 & 1.40650 & -0.19687 & 1.07912 & -0.18946\\ 
& 2 & 1.52239 & -0.41159 & 1.61297 & -0.48358 & 1.40014 & -0.32863 & 1.07133 & -0.31682\\
\hline
\end{tabular}
\caption{Numerical values for $Re(\omega)$ and $Im(\omega)$ for $\alpha=1$}
\label{table:num}
\end{table*}

\begin{table*}[htb!]
\centering
\begin{tabular}{|m{2em}|m{2em}|m{4em}|m{4em}|m{4em}|m{4em}|m{4em}|m{4em}|m{4em}|m{4em}|} 
 \hline
 \multicolumn{2}{|c|}{ } & \multicolumn{2}{|c|}{Model 1}  & \multicolumn{2}{|c|}{Model 2} & \multicolumn{2}{|c|}{Model 3} & \multicolumn{2}{|c|}{Model 4}\\  
 \hline
 L & $n$ & \multicolumn{1}{|c|}{Re $\omega$}  & \multicolumn{1}{|c|}{Im $\omega$} & \multicolumn{1}{|c|}{Re $\omega$}  & \multicolumn{1}{|c|}{Im $\omega$} & \multicolumn{1}{|c|}{Re $\omega$}  & \multicolumn{1}{|c|}{Im $\omega$}& \multicolumn{1}{|c|}{Re $\omega$}  & \multicolumn{1}{|c|}{Im $\omega$}\\
\hline
6 & 0 & 1.09141 & -0.06802 & 1.24568 & -0.09625 & 0.85604 & -0.04576 & 0.62292 & -0.04765\\ 
& 1 & 1.08899 & -0.20429 & 1.23665 & -0.28954 & 0.85263 & -0.13746 & 0.61892 & -0.14334\\ 
& 2 & 1.08457 & -0.34119 & 1.21882 & -0.48523 & 0.84579 & -0.22972 & 0.61110 & -0.24023\\
\hline
7 & 0 & 1.25896 & -0.06801 & 1.43716 & -0.09624 & 0.98766 & -0.04575 & 0.71861 & -0.04764\\ 
& 1 & 1.25682 & -0.20419 & 1.42932 & -0.28931 & 0.98473 & -0.13740 & 0.71514 & -0.14321\\ 
& 2 & 1.25282 & -0.34085 & 1.41379 & -0.48420 & 0.97894 & -0.22943 & 0.70830 & -0.23969\\
\hline
8 & 0 & 1.42654 & -0.06800 & 1.62866 & -0.09623 & 1.11929 & -0.04575 & 0.81431 & -0.04763\\ 
& 1 & 1.42463 & -0.20413 & 1.62173 & -0.28916 & 1.11669 & -0.13736 & 0.81125 & -0.14313\\ 
& 2 & 1.42100 & -0.34063 & 1.60798 & -0.48350 & 1.11151 & -0.22929 & 0.80519 & -0.23933\\
\hline
\end{tabular}
\caption{Numerical values for $Re(\omega)$ and $Im(\omega)$ for $\alpha=2$}
\label{table:num2}
\end{table*}

\begin{table*}[htb!]
\centering
\begin{tabular}{|m{2em}|m{2em}|m{4em}|m{4em}|m{4em}|m{4em}|m{4em}|m{4em}|m{4em}|m{4em}|} 
 \hline
 \multicolumn{2}{|c|}{ } & \multicolumn{2}{|c|}{Model 1}  & \multicolumn{2}{|c|}{Model 2} & \multicolumn{2}{|c|}{Model 3} & \multicolumn{2}{|c|}{Model 4}\\  
 \hline
 L & $n$ & \multicolumn{1}{|c|}{Re $\omega$}  & \multicolumn{1}{|c|}{Im $\omega$} & \multicolumn{1}{|c|}{Re $\omega$}  & \multicolumn{1}{|c|}{Im $\omega$} & \multicolumn{1}{|c|}{Re $\omega$}  & \multicolumn{1}{|c|}{Im $\omega$}& \multicolumn{1}{|c|}{Re $\omega$}  & \multicolumn{1}{|c|}{Im $\omega$}\\
\hline
6 & 0 & 1.00397 & -0.05442 & 1.24261 & -0.09624 & 0.65548 & -0.03756 & 0.49957 & -0.03835\\ 
& 1 & 1.00447 & -0.16356 & 1.23312 & -0.28953 & 0.65348 & -0.11284 & 0.49631 & -0.11536\\ 
& 2 & 1.00574 & -0.27374 & 1.21432 & -0.48523 & 0.64954 & -0.18853 & 0.48993 & -0.19333\\
\hline
7 & 0 & 1.15798 & -0.05439 & 1.43368 & -0.09624 & 0.75619 & -0.03756 & 0.57632 & -0.03834\\ 
& 1 & 1.15839 & -0.16340 & 1.42545 & -0.28932 & 0.75445 & -0.11279 & 0.57348 & -0.11526\\ 
& 2 & 1.15941 & -0.27313 & 1.40909 & -0.48421 & 0.75102 & -0.18834 & 0.56791 & -0.19290\\
\hline
8 & 0 & 1.31204 & -0.05438 & 1.62476 & -0.09623 & 0.85692 & -0.03756 & 0.65308 & -0.03834\\ 
& 1 & 1.31238 & -0.16330 & 1.61749 & -0.28917 & 0.85538 & -0.11276 & 0.65057 & -0.11520\\ 
& 2 & 1.31322 & -0.27275 & 1.60303 & -0.48353 & 0.85235 & -0.18820 & 0.64563 & -0.19262\\
\hline
\end{tabular}
\caption{Numerical values for $Re(\omega)$ and $Im(\omega)$ for $\alpha=3$}
\label{table:num3}
\end{table*}

\begin{figure*}[hbt!]
\centering
\includegraphics[width=0.27\textwidth]{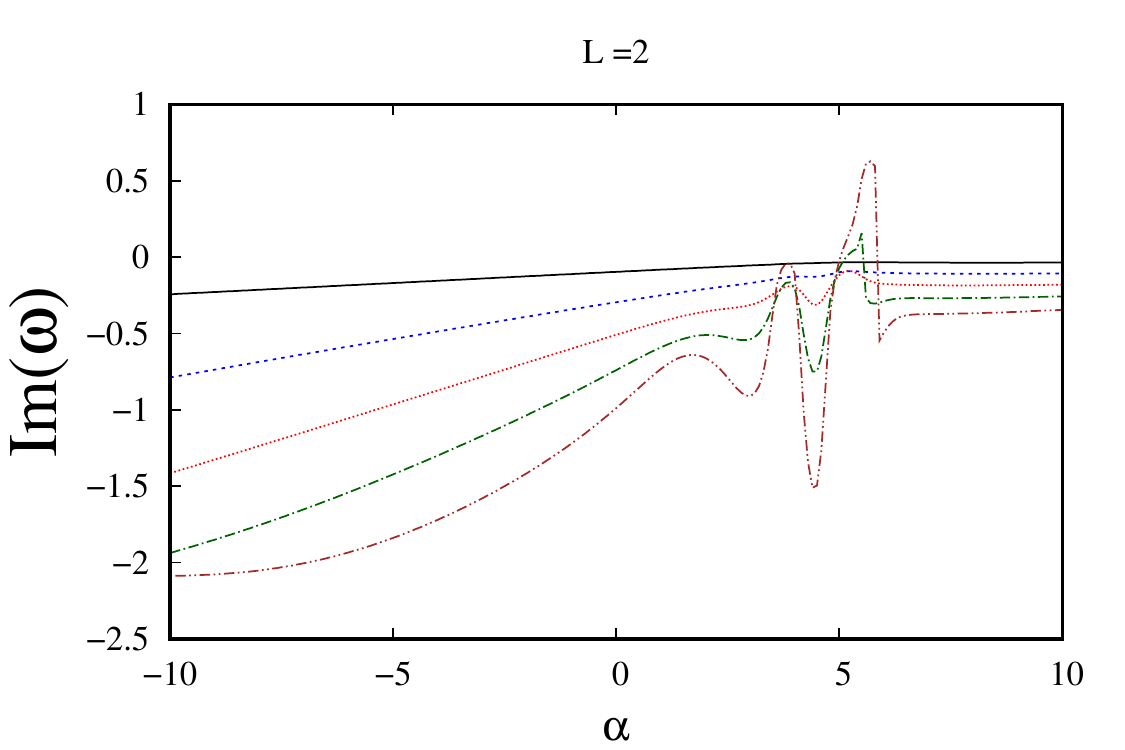}  \
\includegraphics[width=0.27\textwidth]{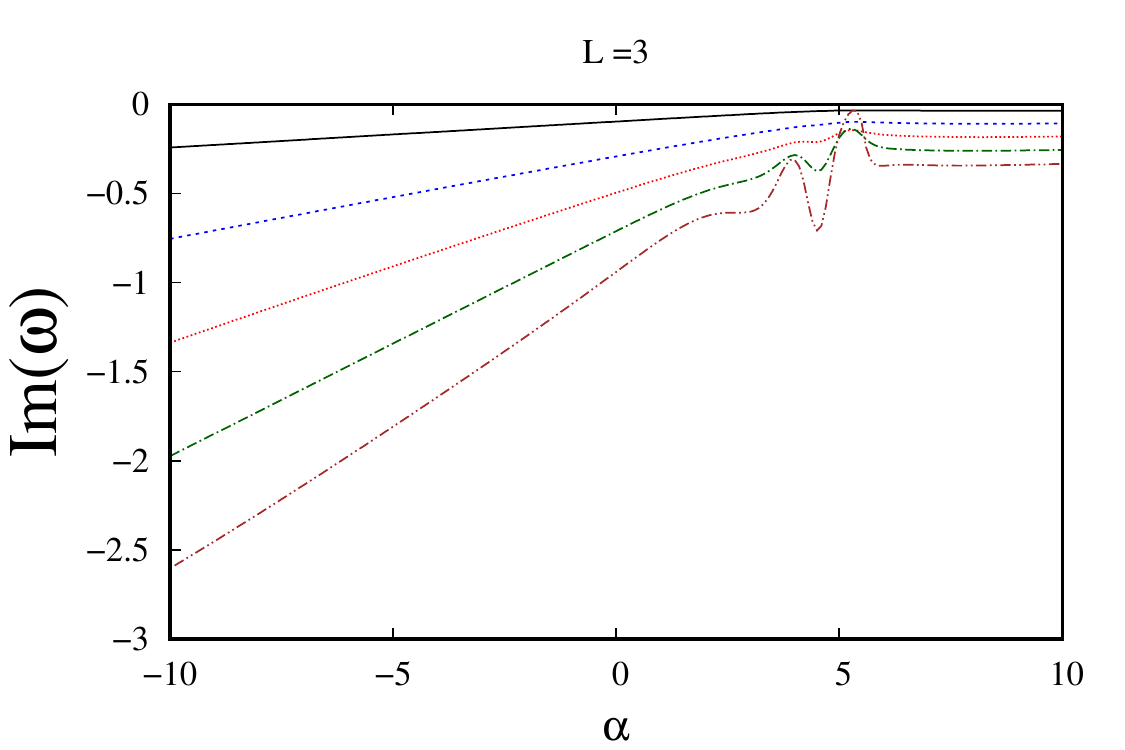}  \
\includegraphics[width=0.27\textwidth]{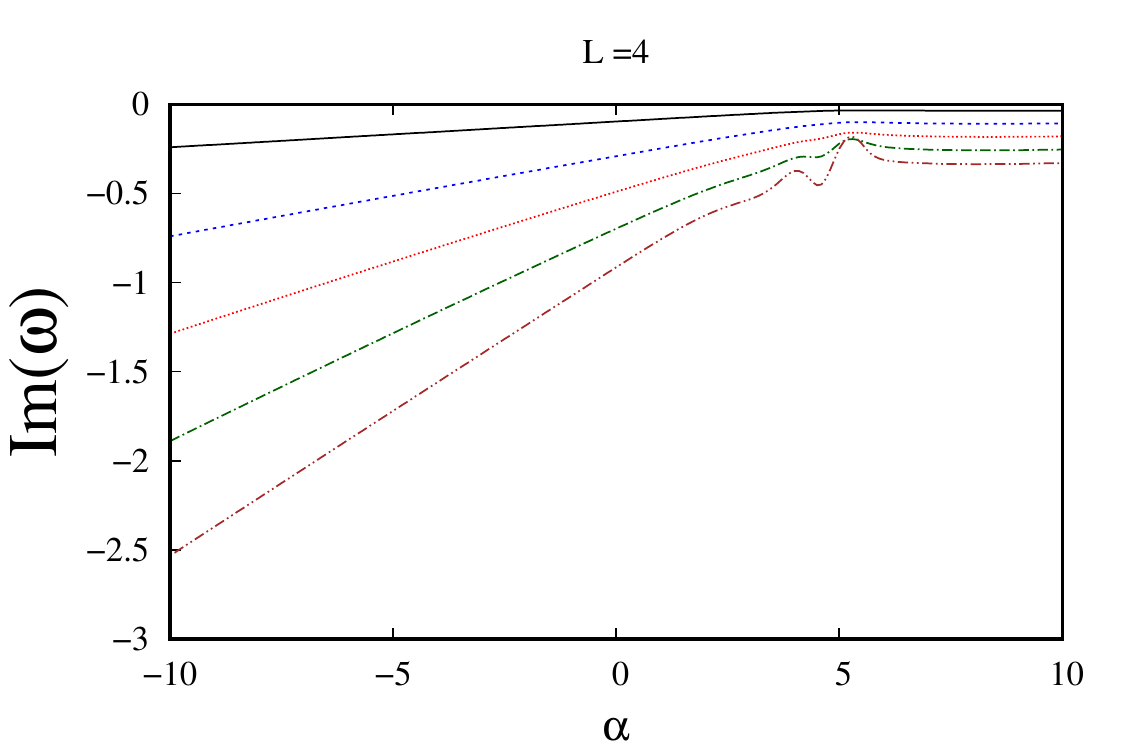}  \
\includegraphics[width=0.27\textwidth]{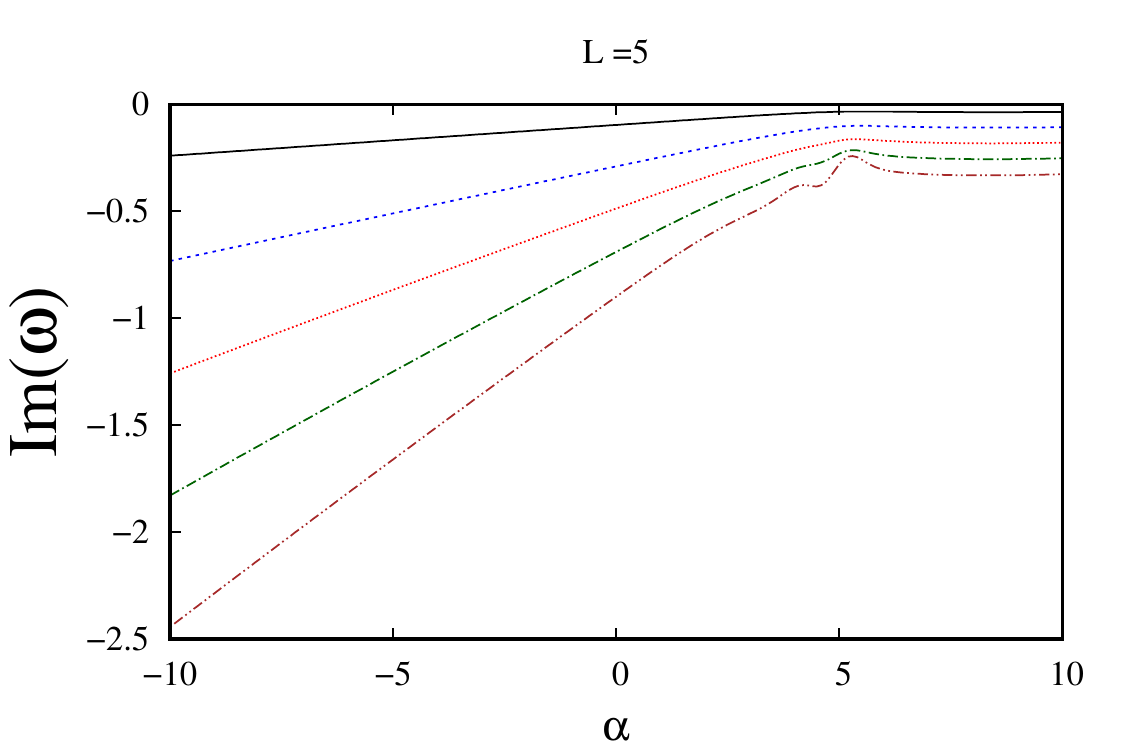}  \
\includegraphics[width=0.27\textwidth]{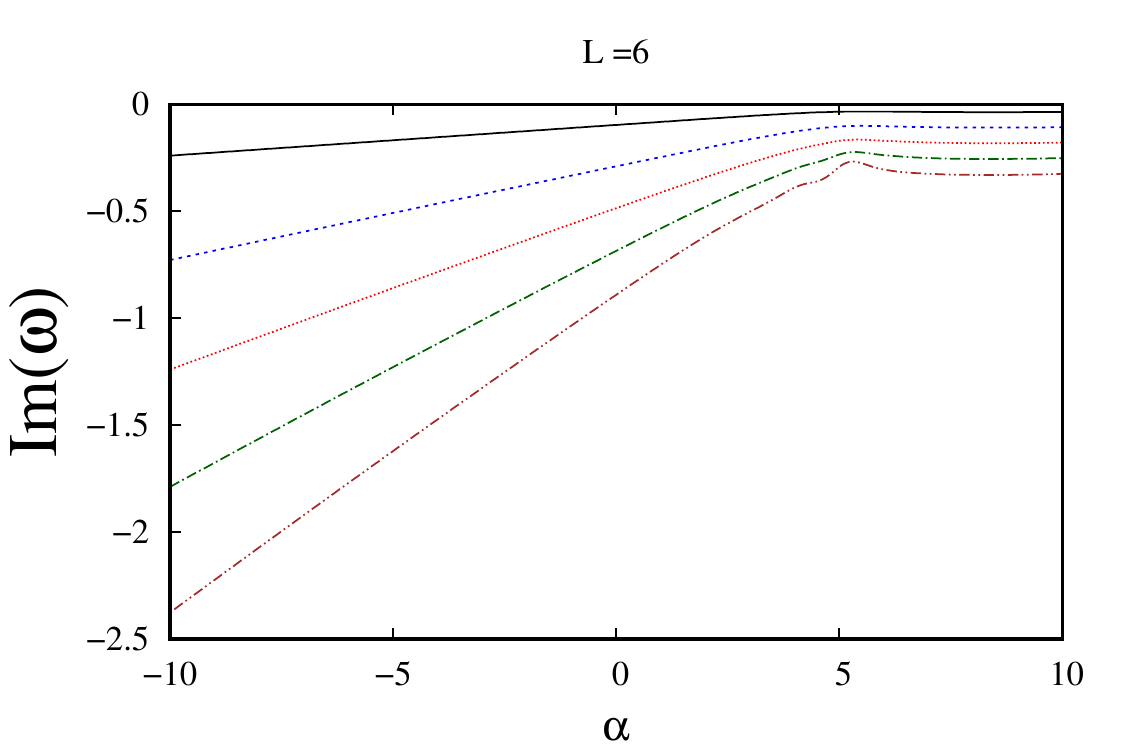}  \
\includegraphics[width=0.27\textwidth]{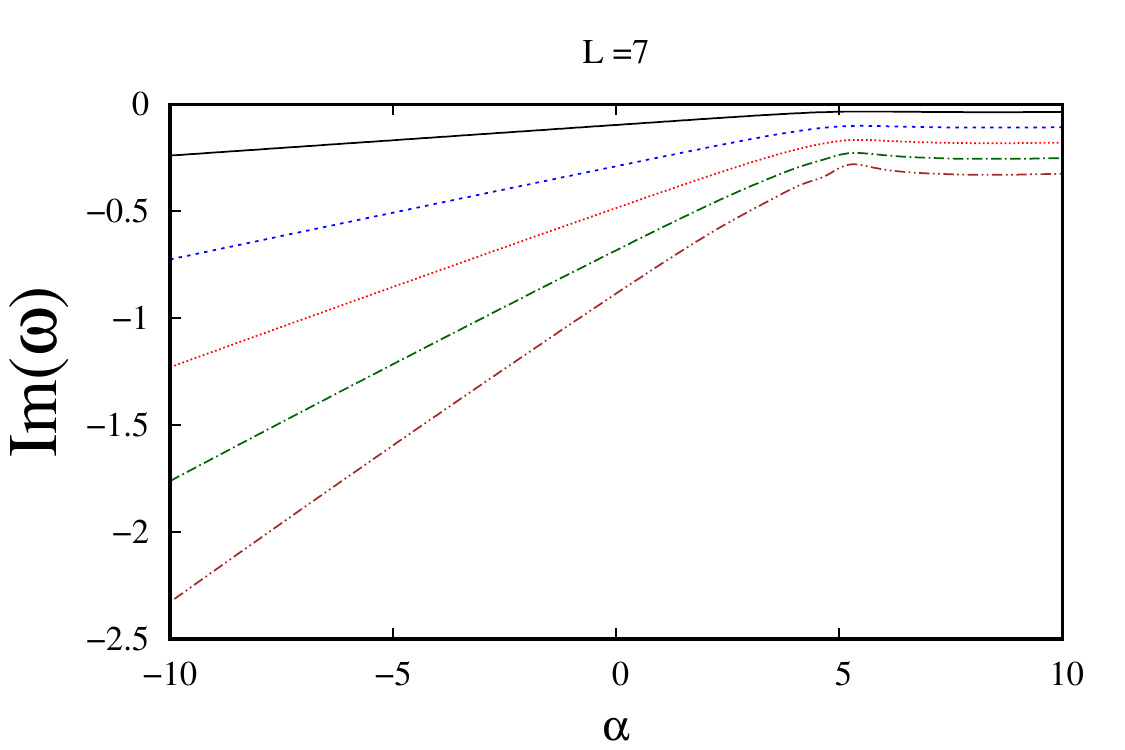}  \
\includegraphics[width=0.27\textwidth]{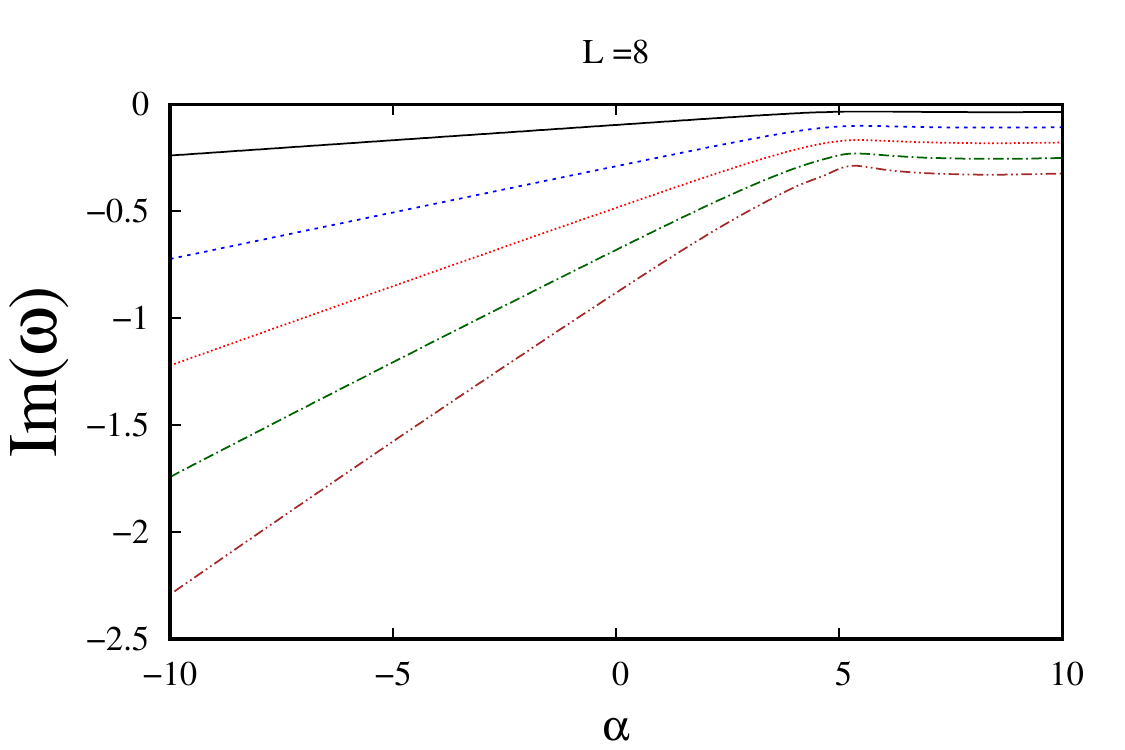}  \
\includegraphics[width=0.27\textwidth]{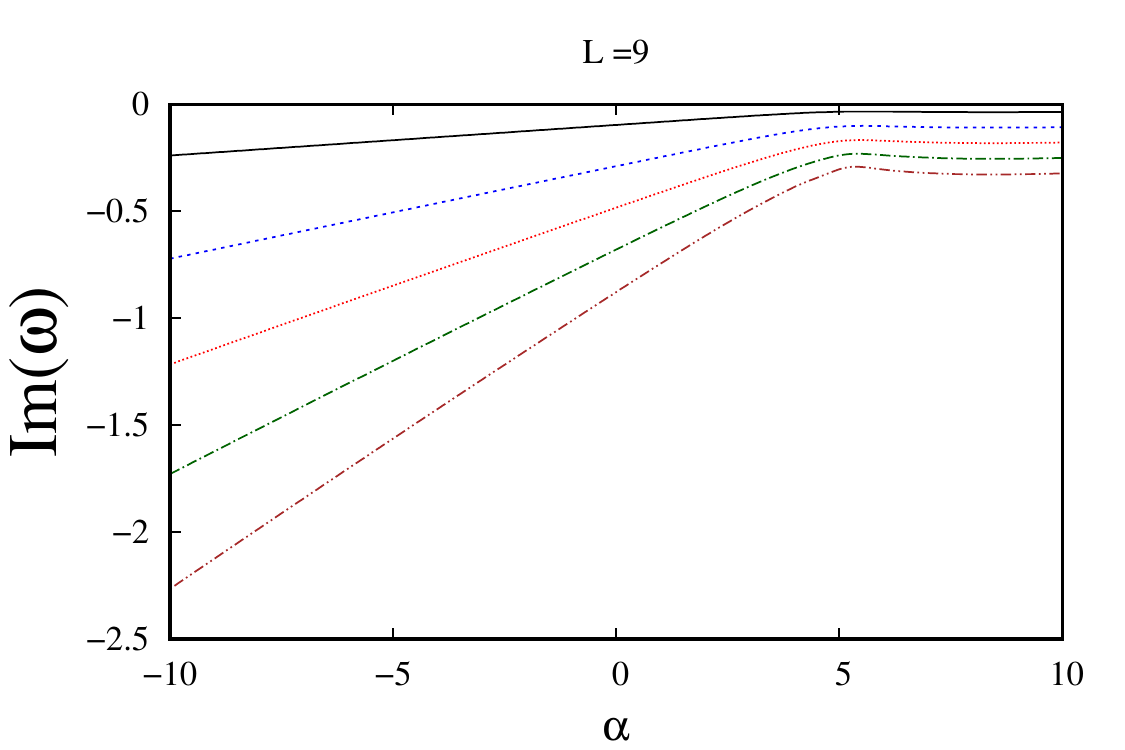}
\includegraphics[width=0.27\textwidth]{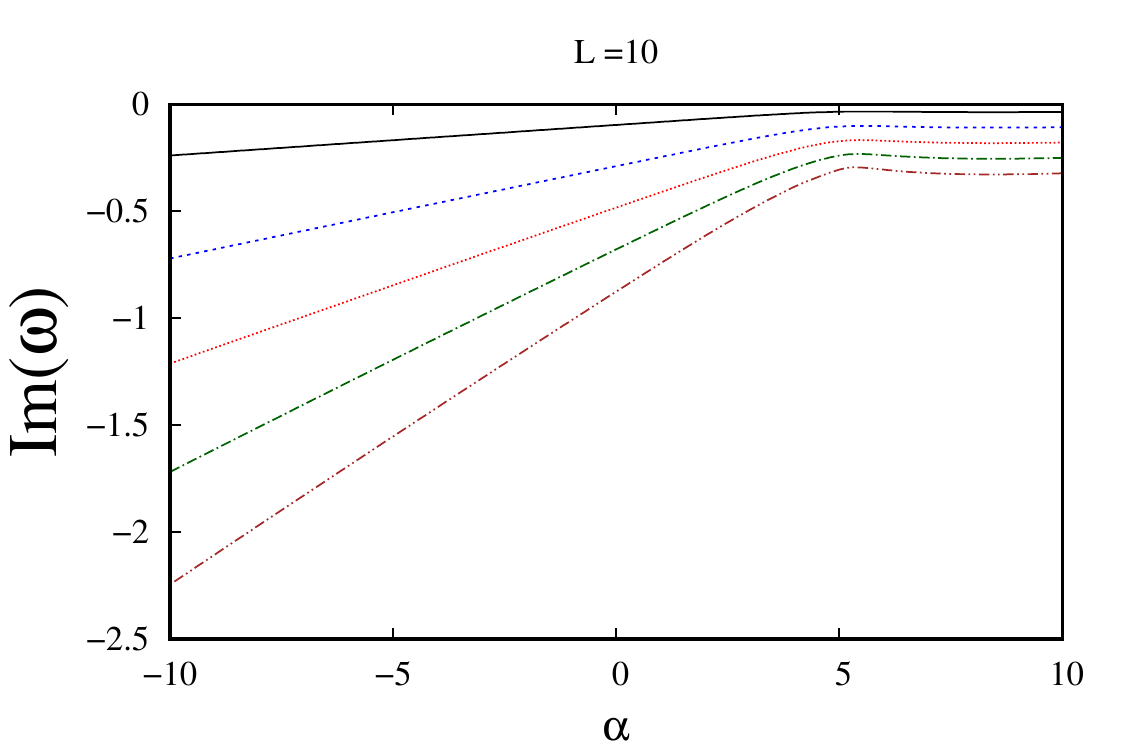}  \
\caption{\label{M1}
Imaginary part of the QNM for Model 1 as a function of the hair $\alpha$ for different values of $L$ and $n$. Each plot corresponds to a different value of $L$. The values for $n$ are $0$ (black line), $1$ (blue line), $2$ (red line), $3$ (green line), $4$ (brown line). }
\end{figure*}

\begin{figure*}[hbt!]
\centering
\includegraphics[width=0.27\textwidth]{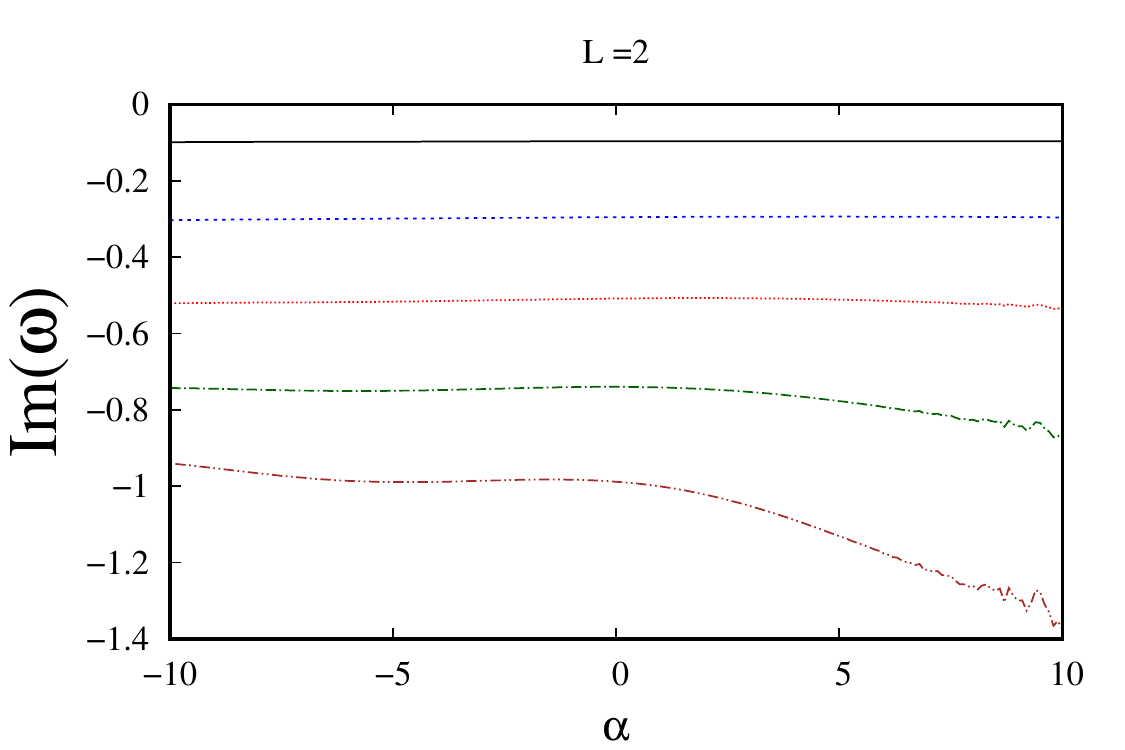}  \
\includegraphics[width=0.27\textwidth]{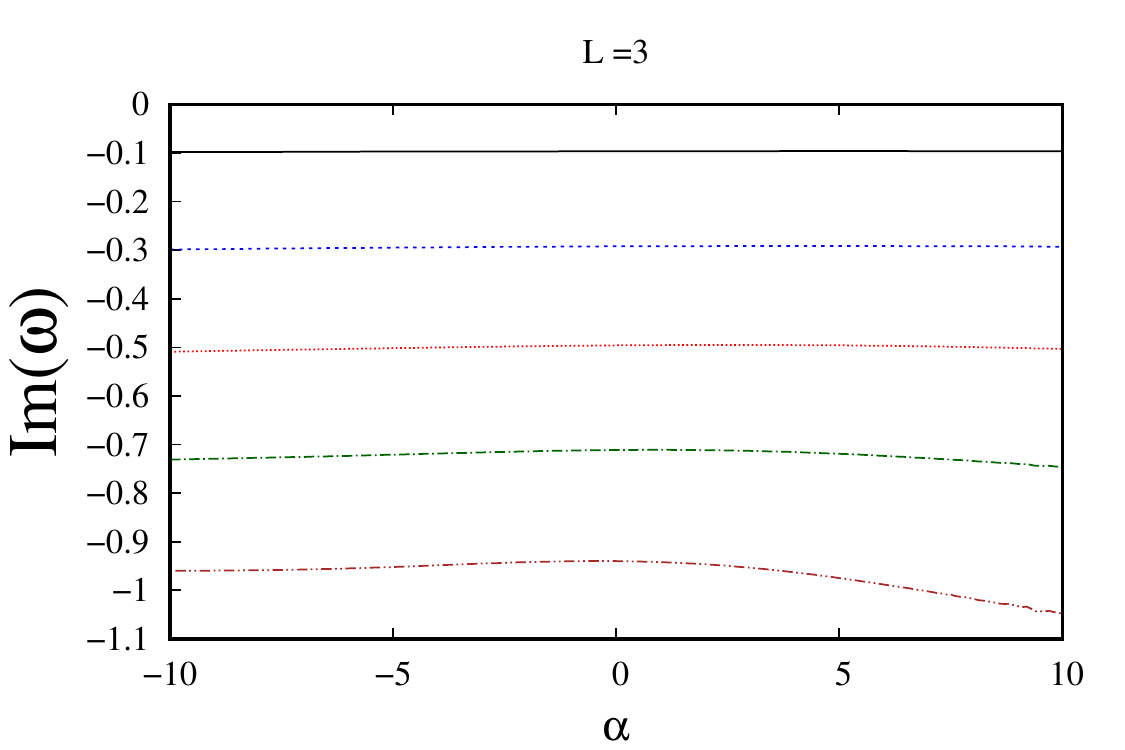}  \
\includegraphics[width=0.27\textwidth]{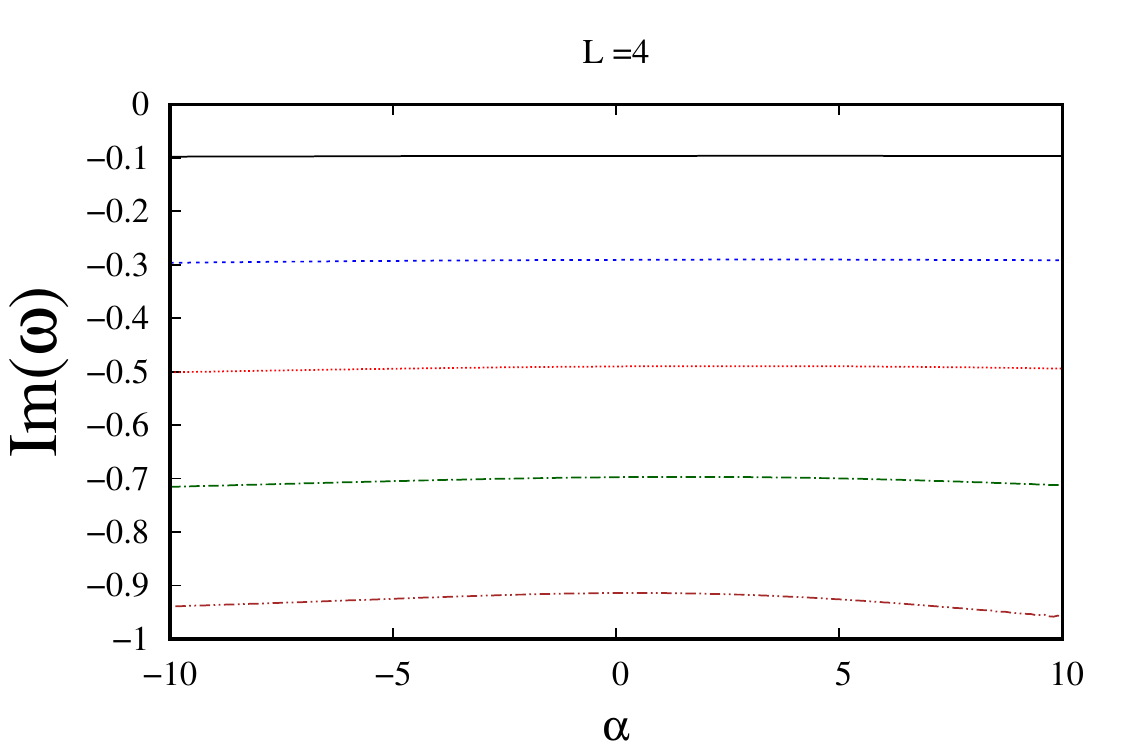}  \
\includegraphics[width=0.27\textwidth]{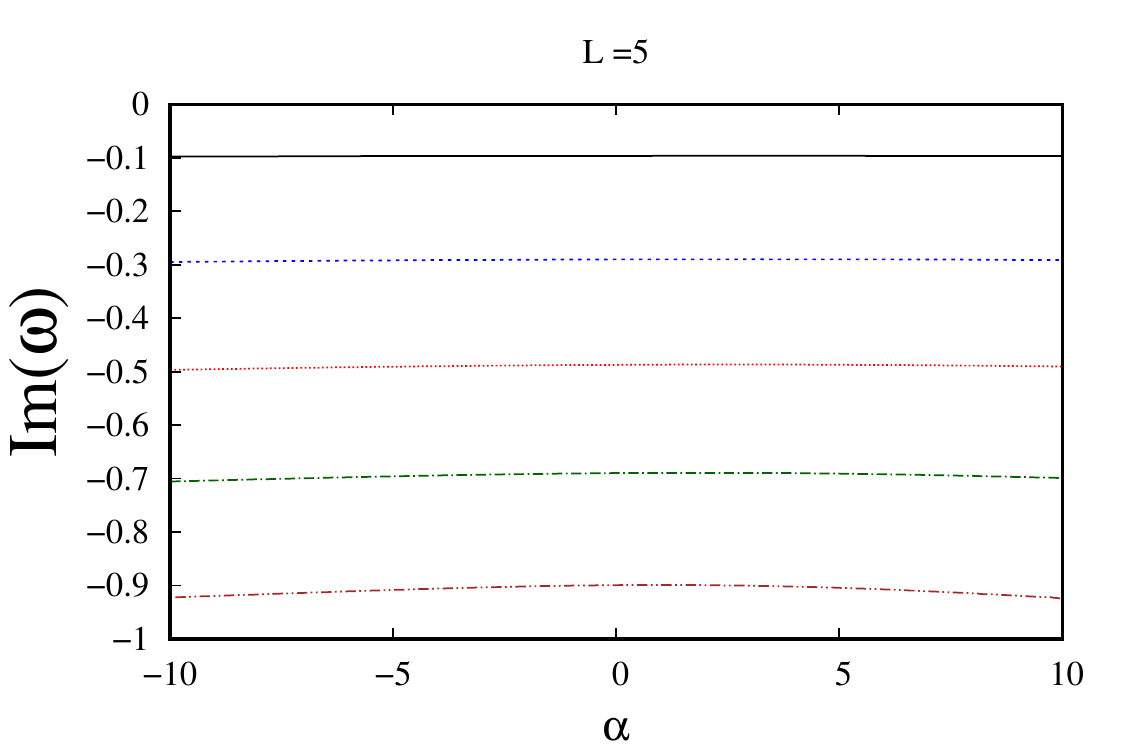}  \
\includegraphics[width=0.27\textwidth]{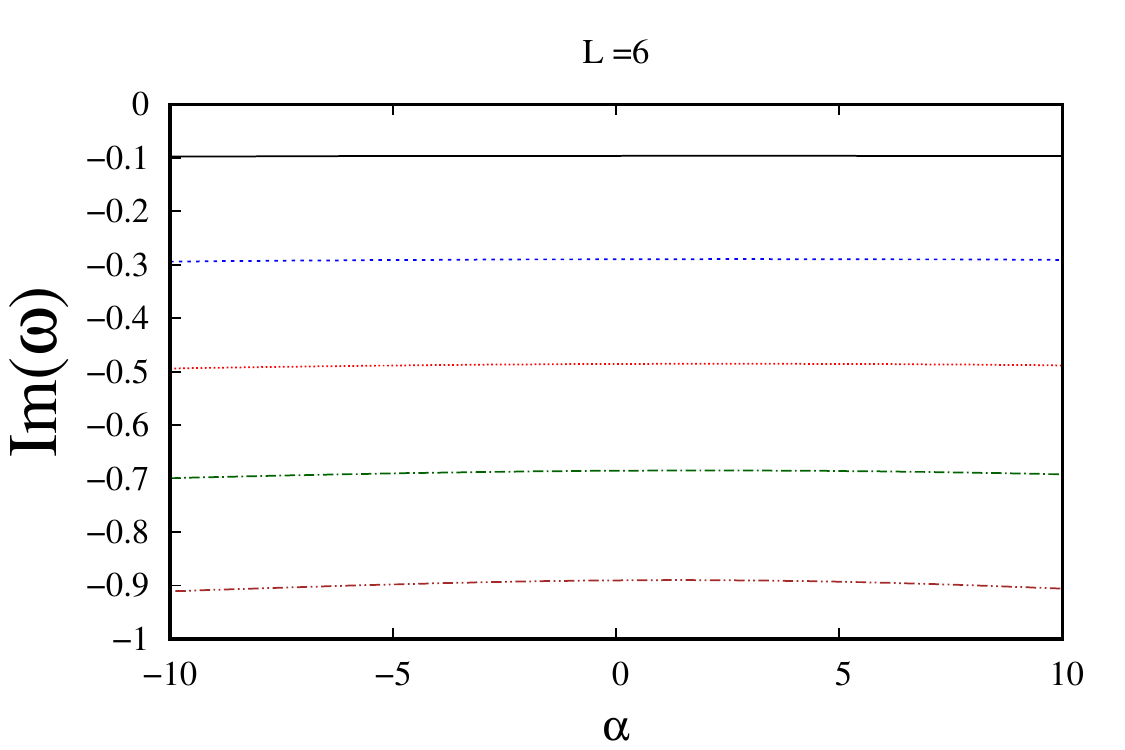}  \
\includegraphics[width=0.27\textwidth]{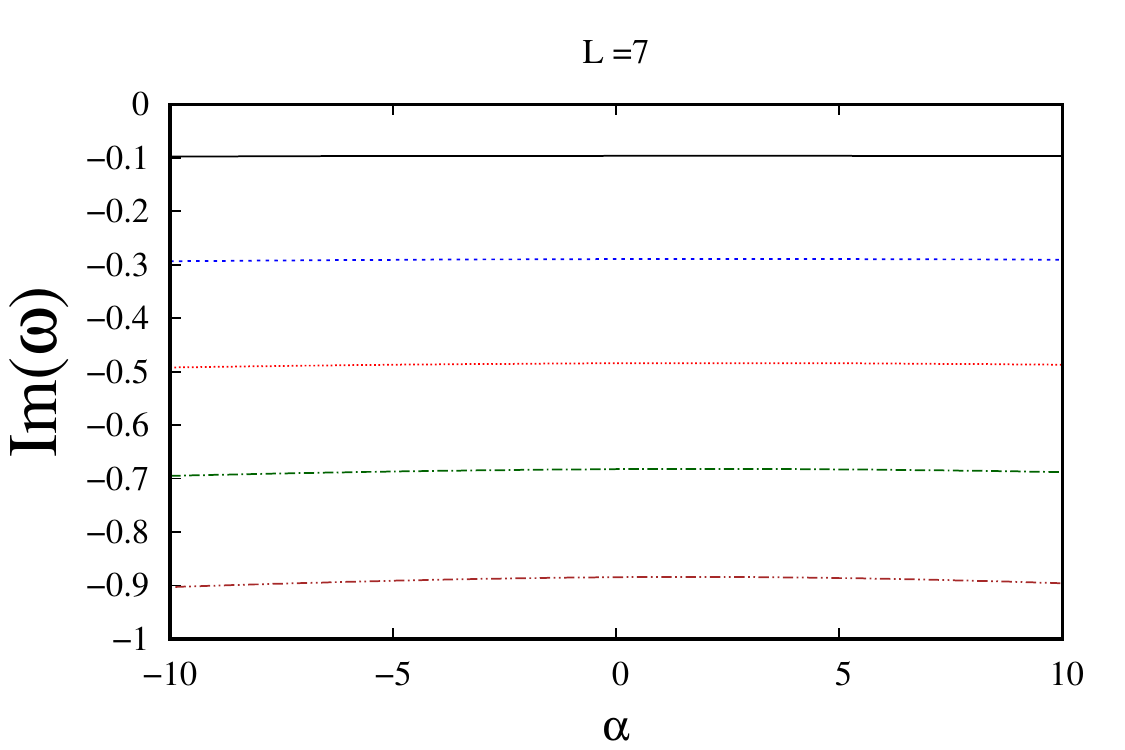}  \
\includegraphics[width=0.27\textwidth]{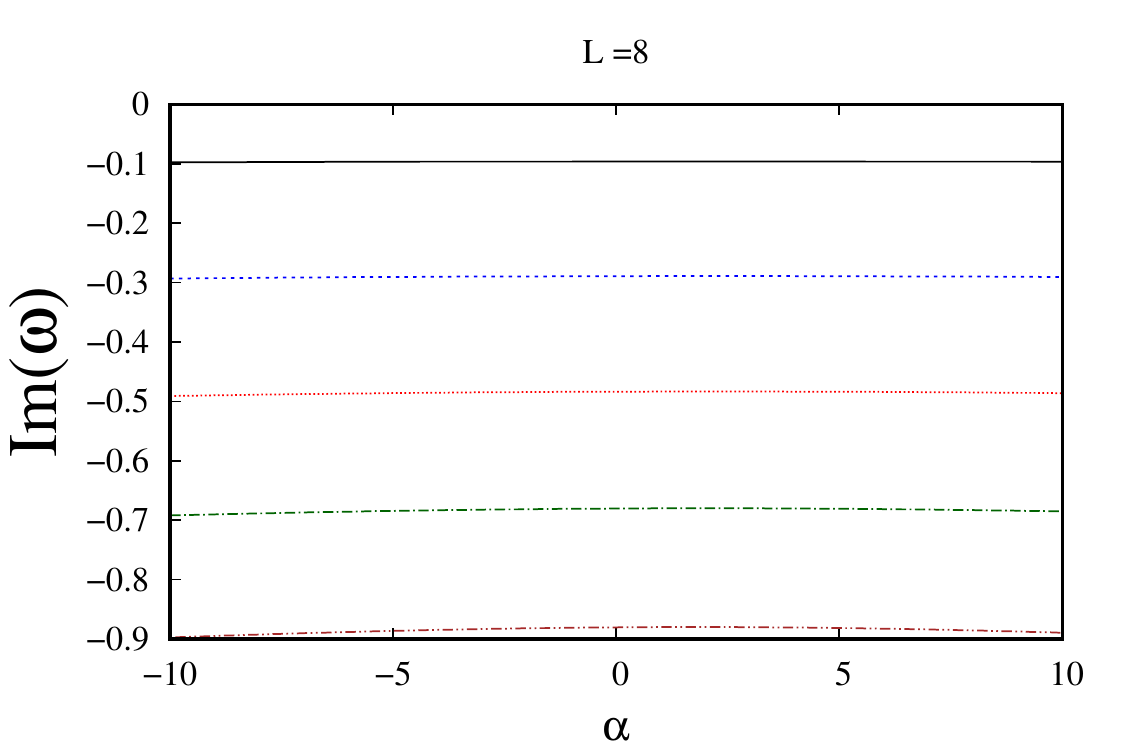}  \
\includegraphics[width=0.27\textwidth]{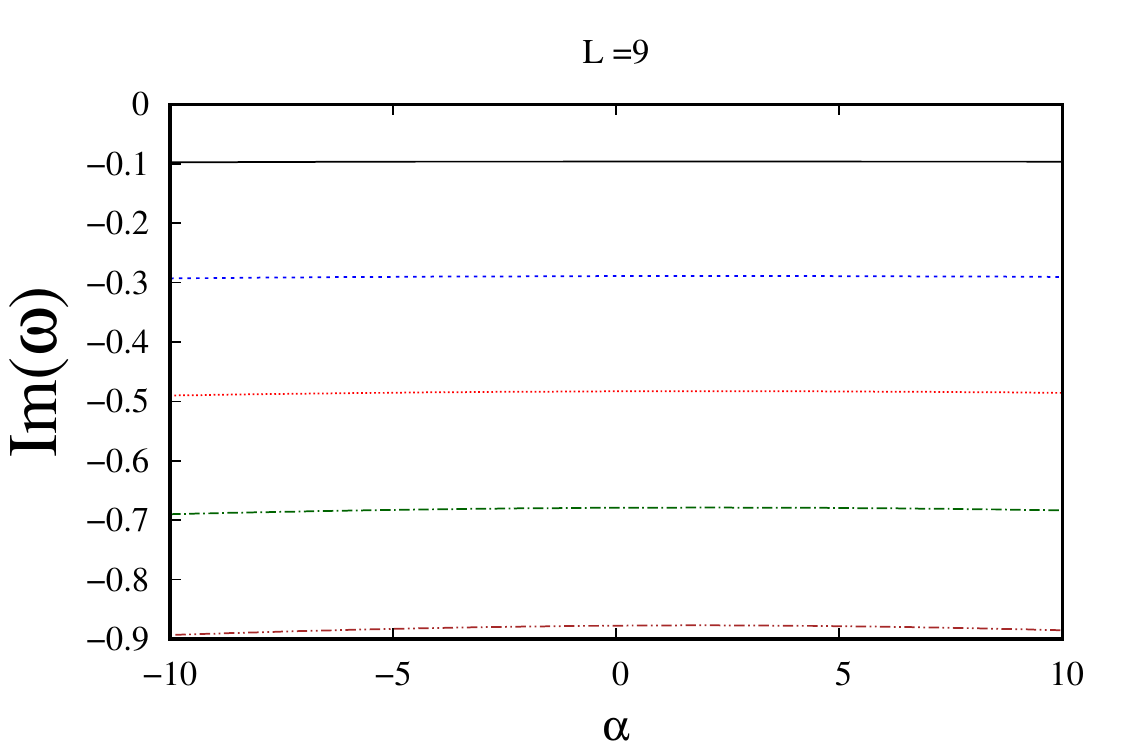}
\includegraphics[width=0.27\textwidth]{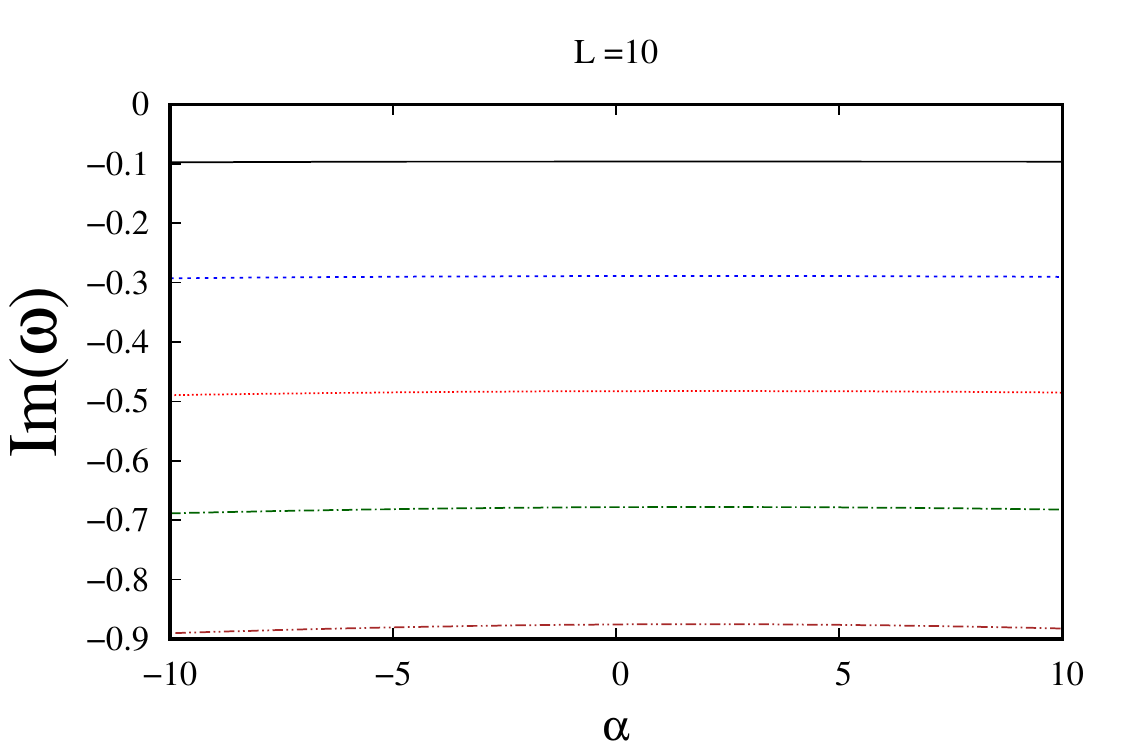}  \
\caption{\label{M2}
Imaginary part of the QNM for Model 2 as a function of the hair $\alpha$ for different values of $L$ and $n$. Each plot corresponds to a different value of $L$. The values for $n$ are $0$ (black line), $1$ (blue line), $2$ (red line), $3$ (green line), $4$ (brown line). }
\end{figure*}

\begin{figure*}[hbt!]
\centering
\includegraphics[width=0.27\textwidth]{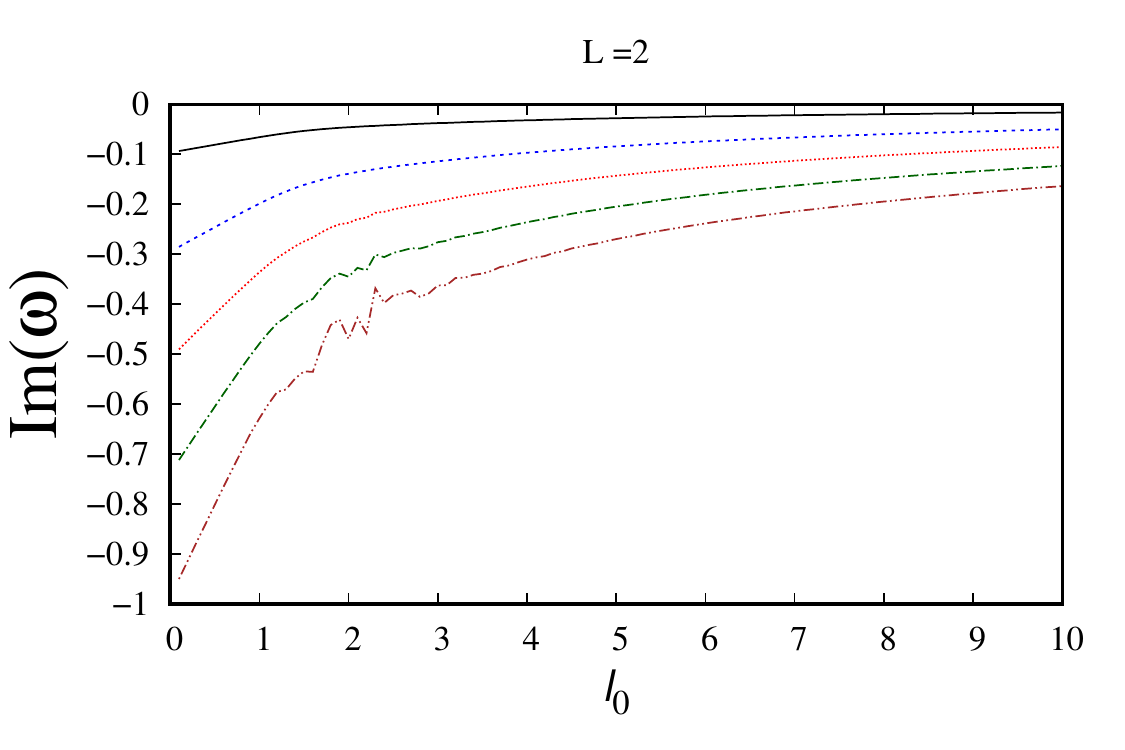}  \
\includegraphics[width=0.27\textwidth]{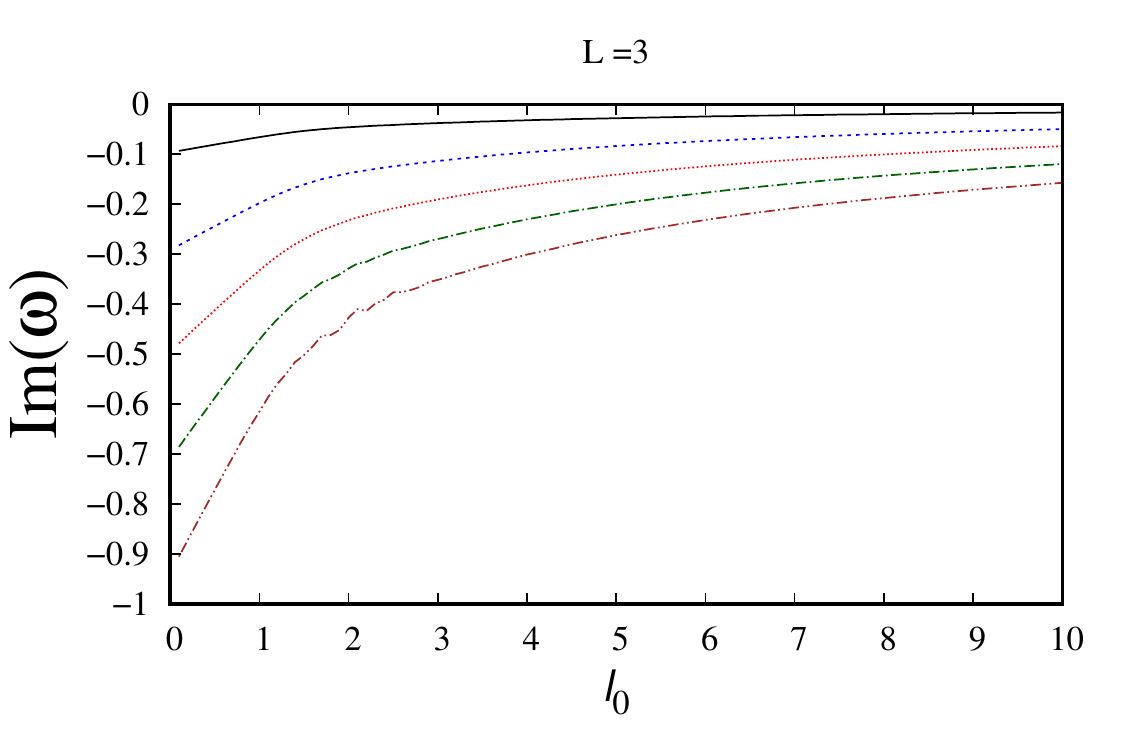}  \
\includegraphics[width=0.27\textwidth]{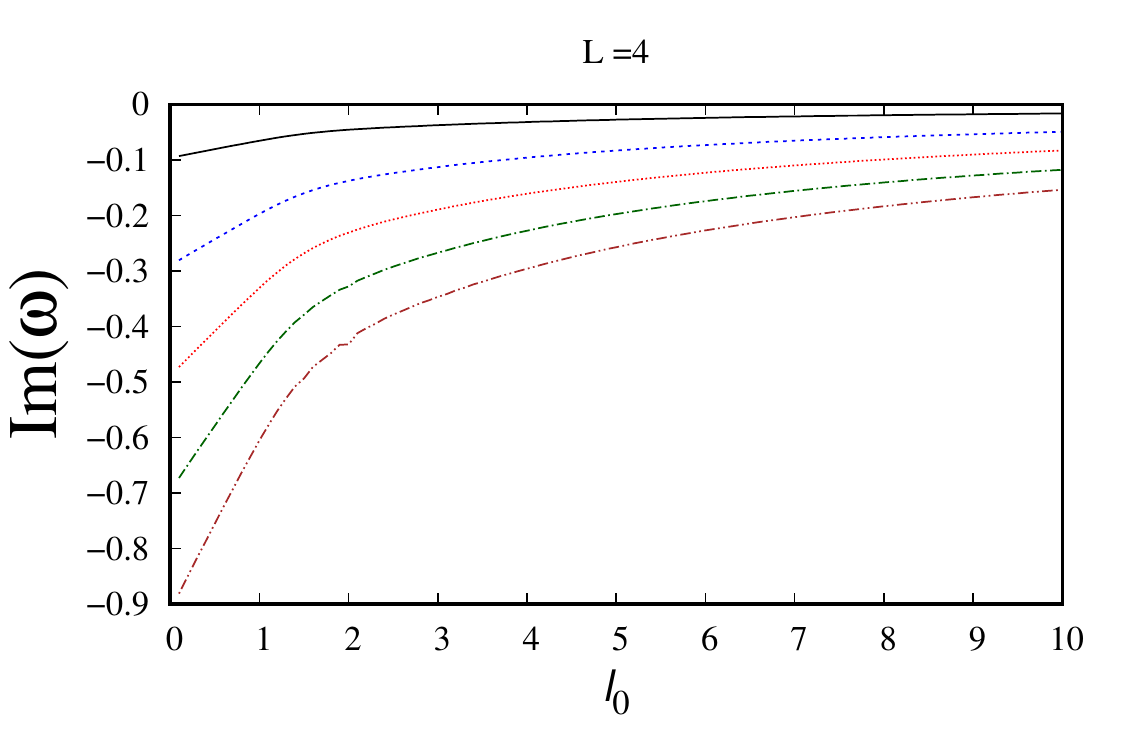}  \
\includegraphics[width=0.27\textwidth]{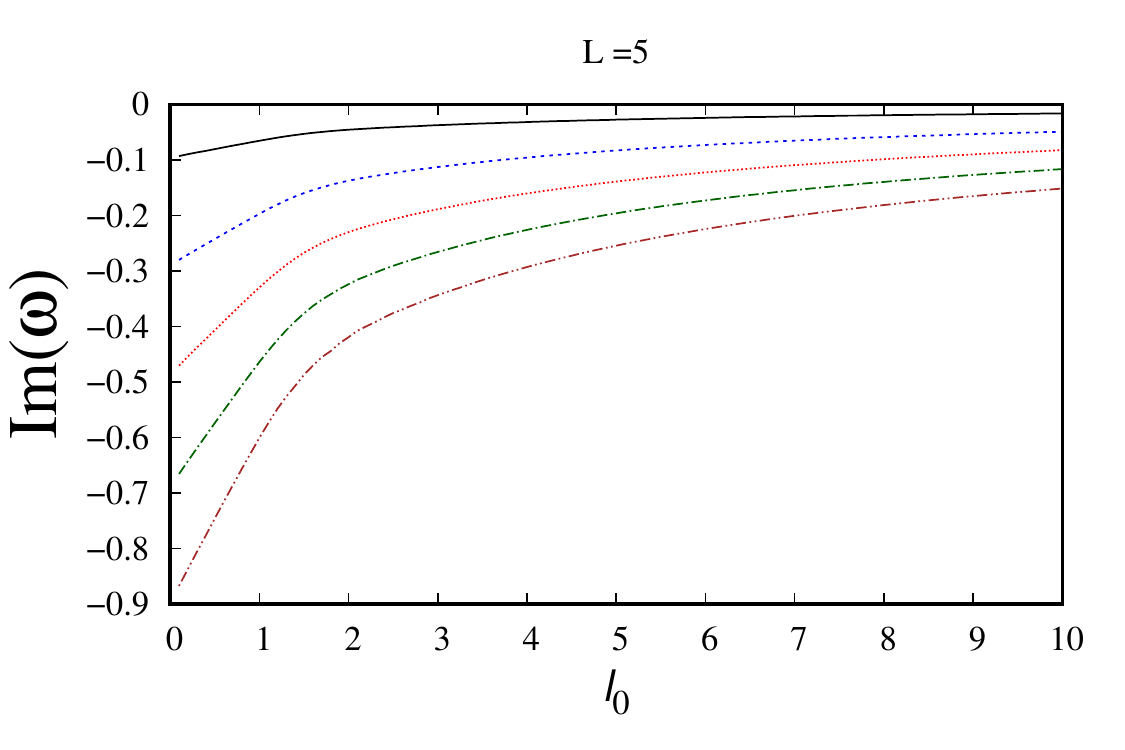}  \
\includegraphics[width=0.27\textwidth]{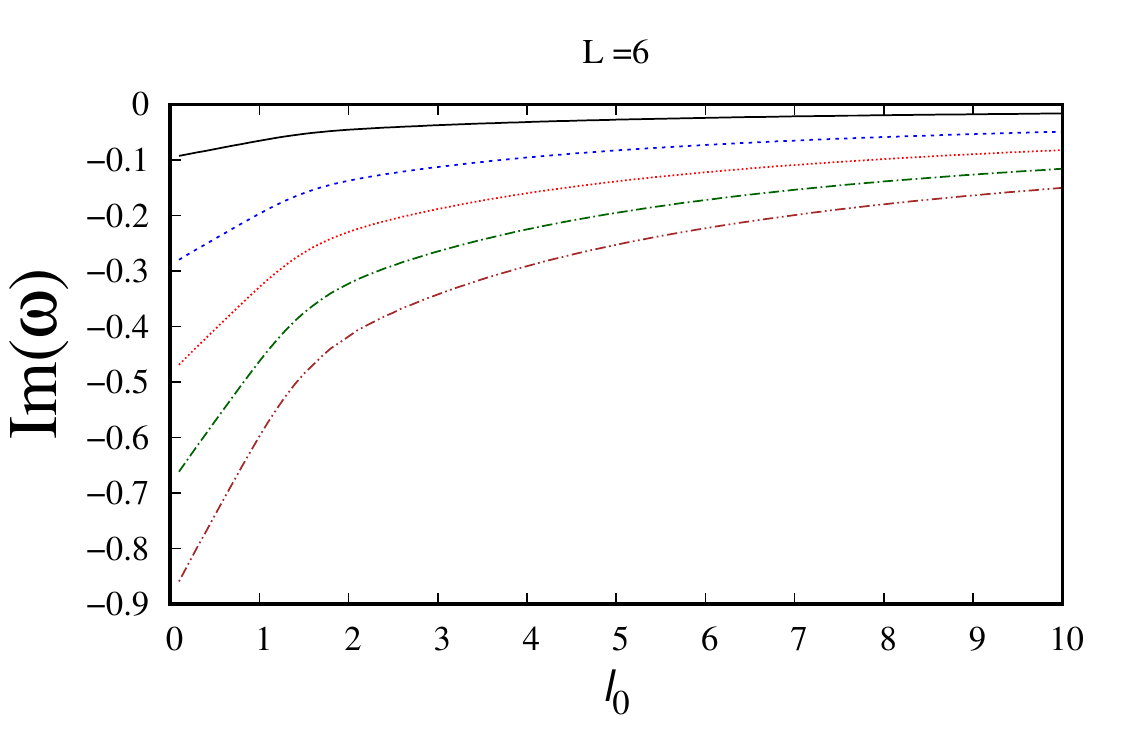}  \
\includegraphics[width=0.27\textwidth]{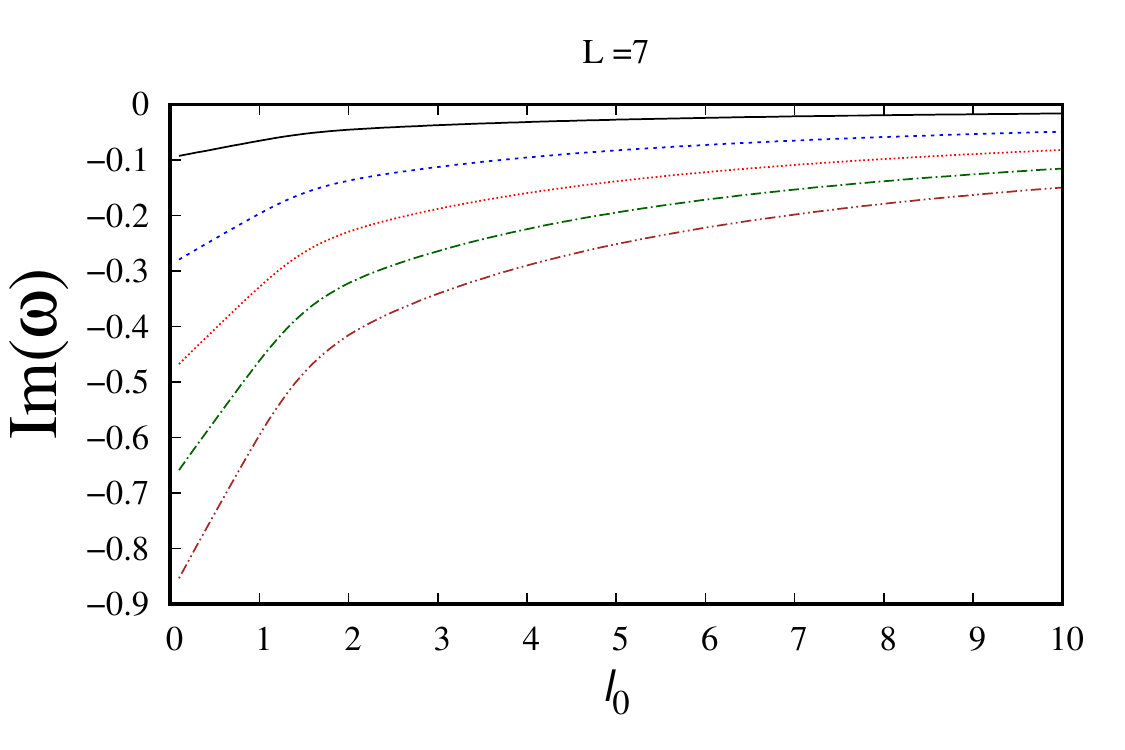}  \
\includegraphics[width=0.27\textwidth]{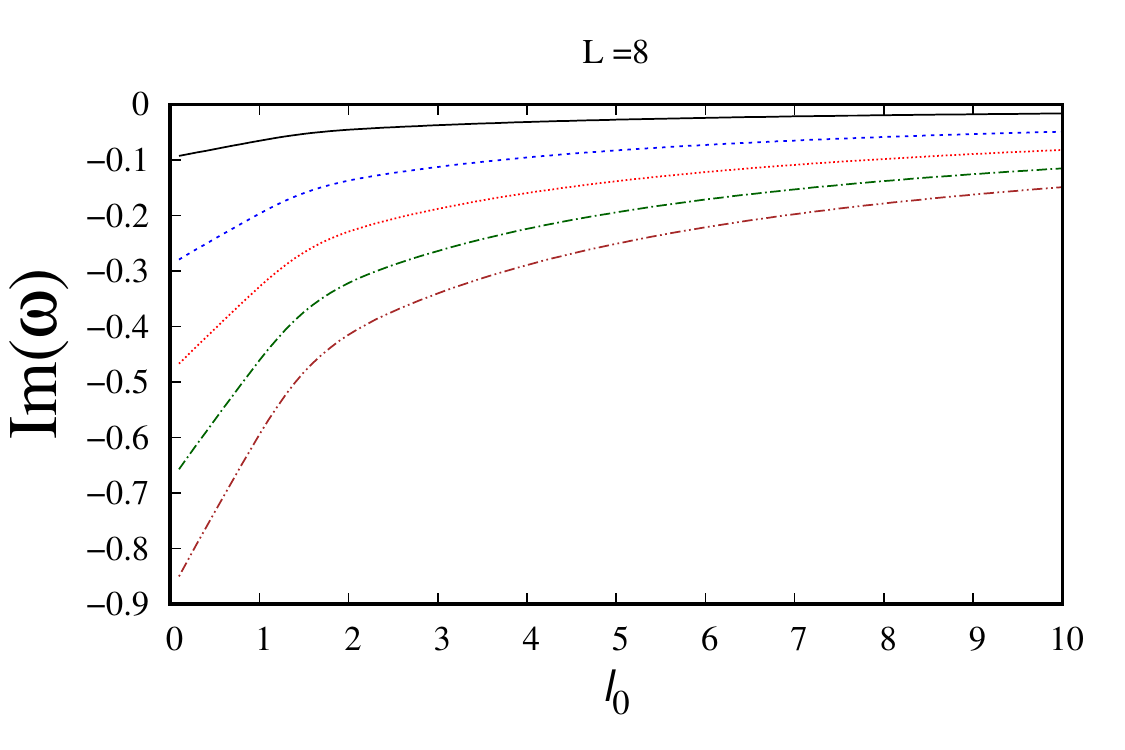}  \
\includegraphics[width=0.27\textwidth]{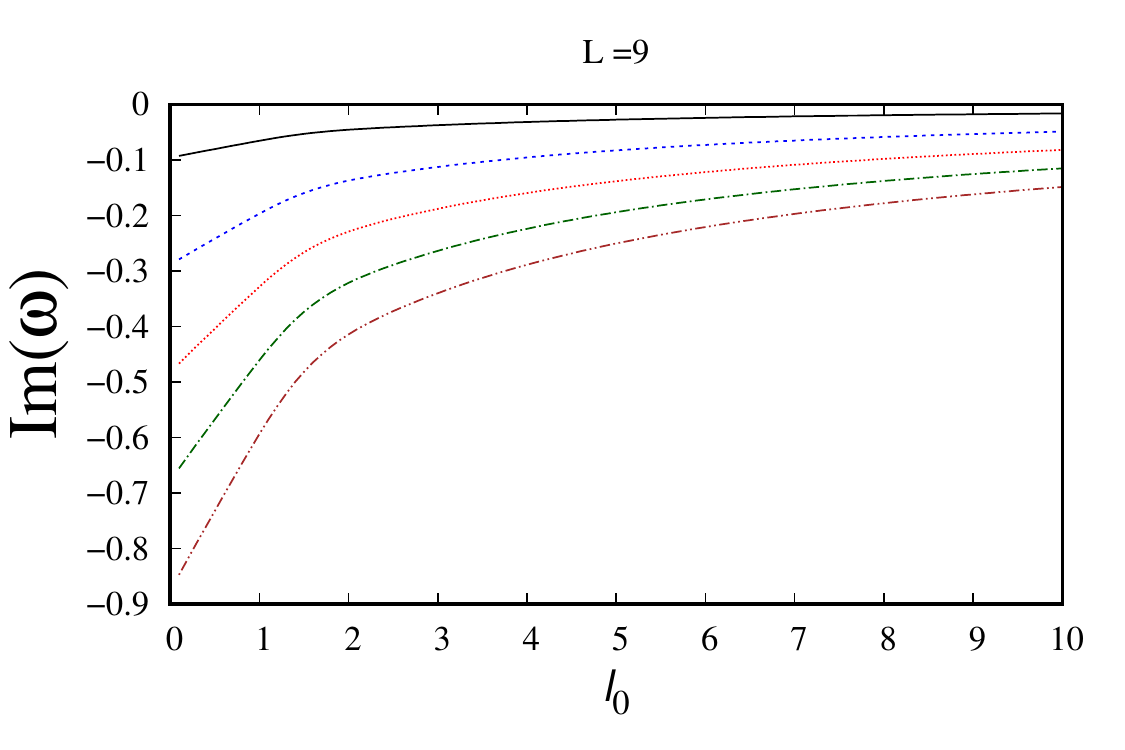}
\includegraphics[width=0.27\textwidth]{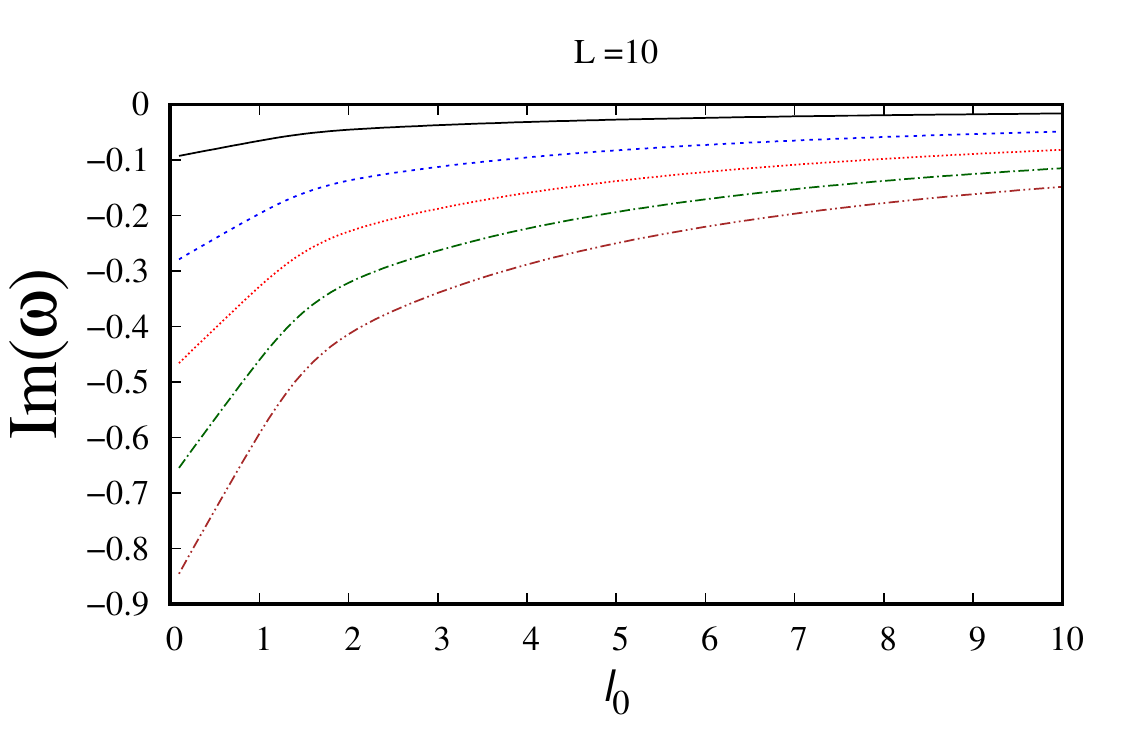}  \
\caption{\label{M3}
Imaginary part of the QNM for Model 3 as a function of the hair $\alpha$ for different values of $L$ and $n$. Each plot corresponds to a different value of $L$. The values for $n$ are $0$ (black line), $1$ (blue line), $2$ (red line), $3$ (green line), $4$ (brown line). }
\end{figure*}

\begin{figure*}[hbt!]
\centering
\includegraphics[width=0.27\textwidth]{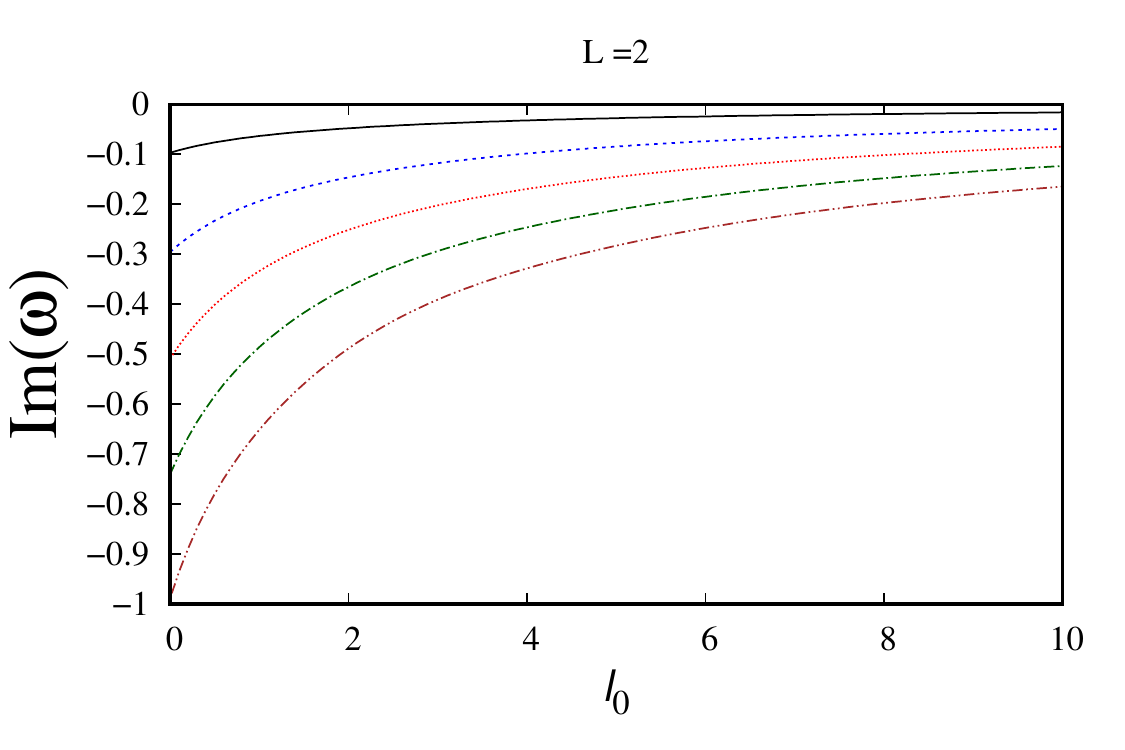}  \
\includegraphics[width=0.27\textwidth]{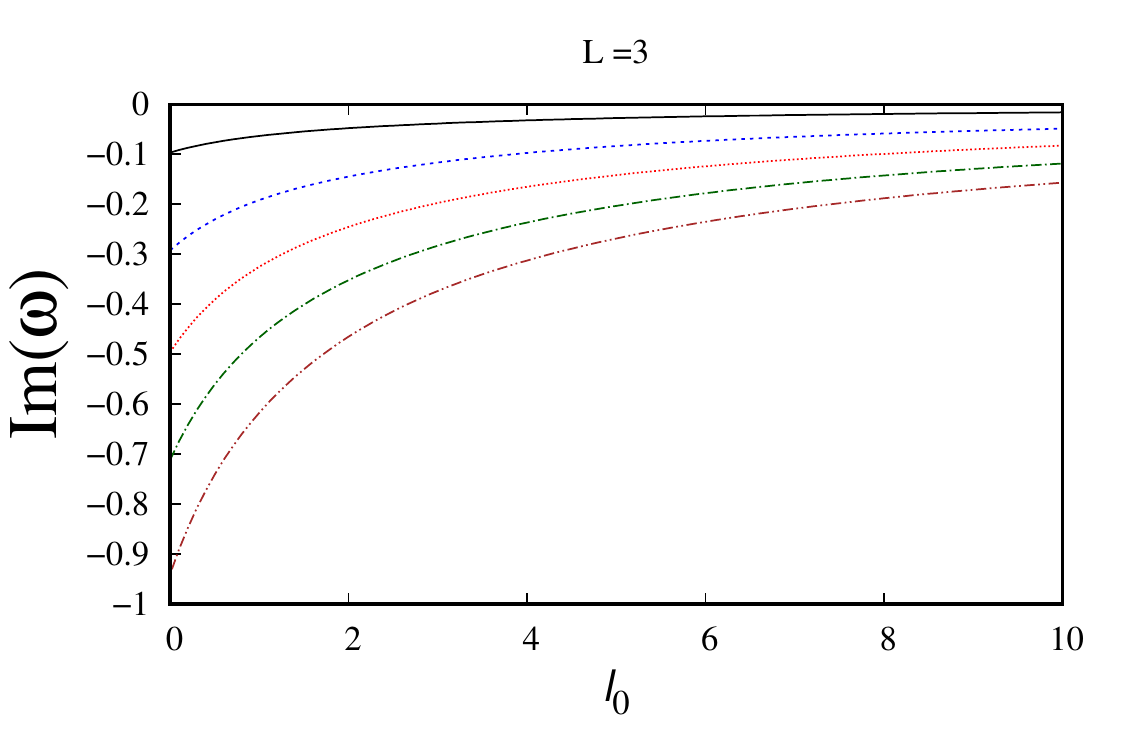}  \
\includegraphics[width=0.27\textwidth]{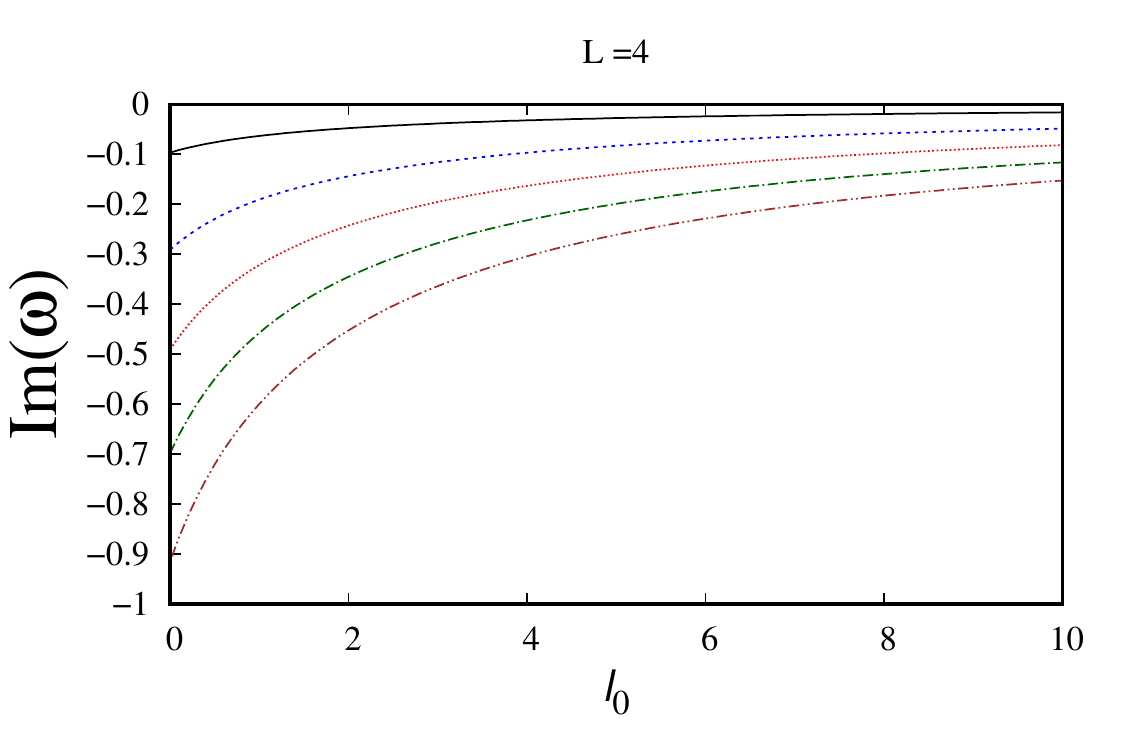}  \
\includegraphics[width=0.27\textwidth]{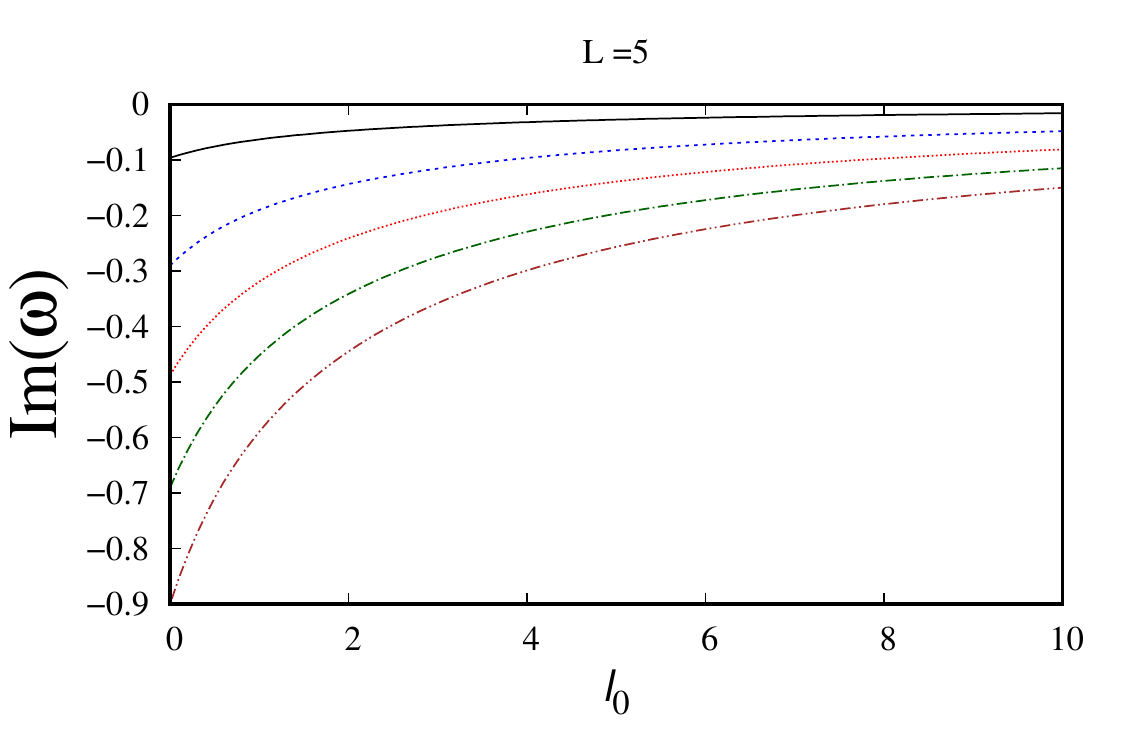}  \
\includegraphics[width=0.27\textwidth]{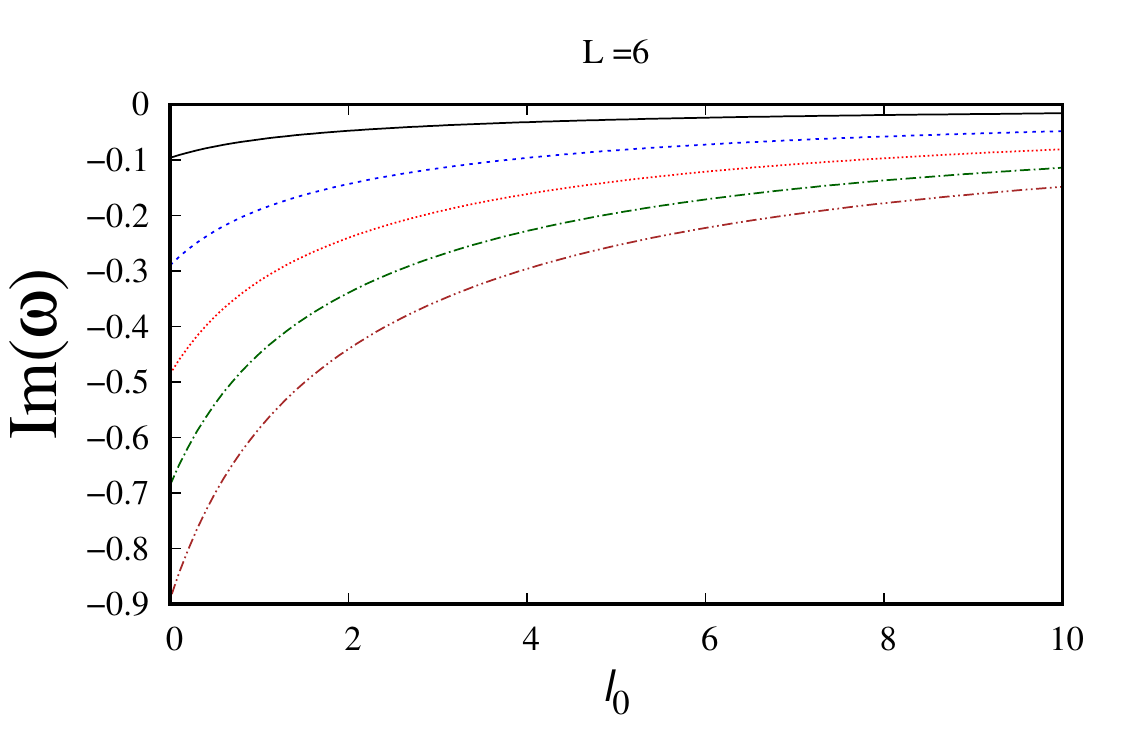}  \
\includegraphics[width=0.27\textwidth]{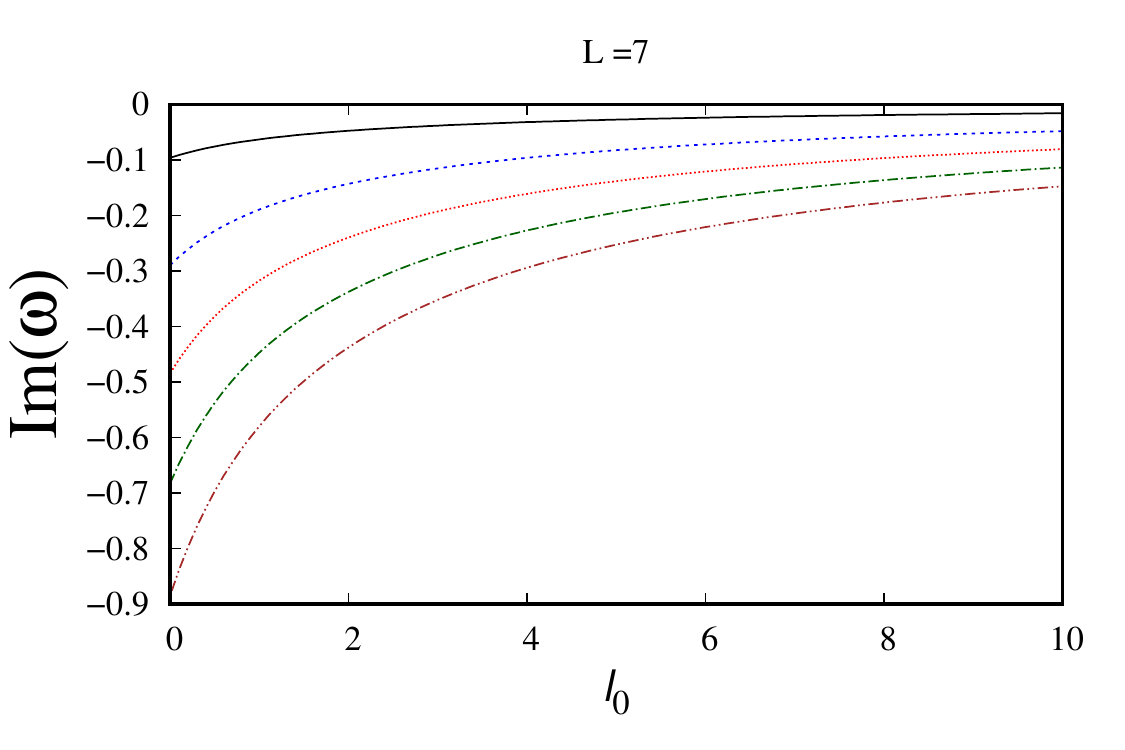}  \
\includegraphics[width=0.27\textwidth]{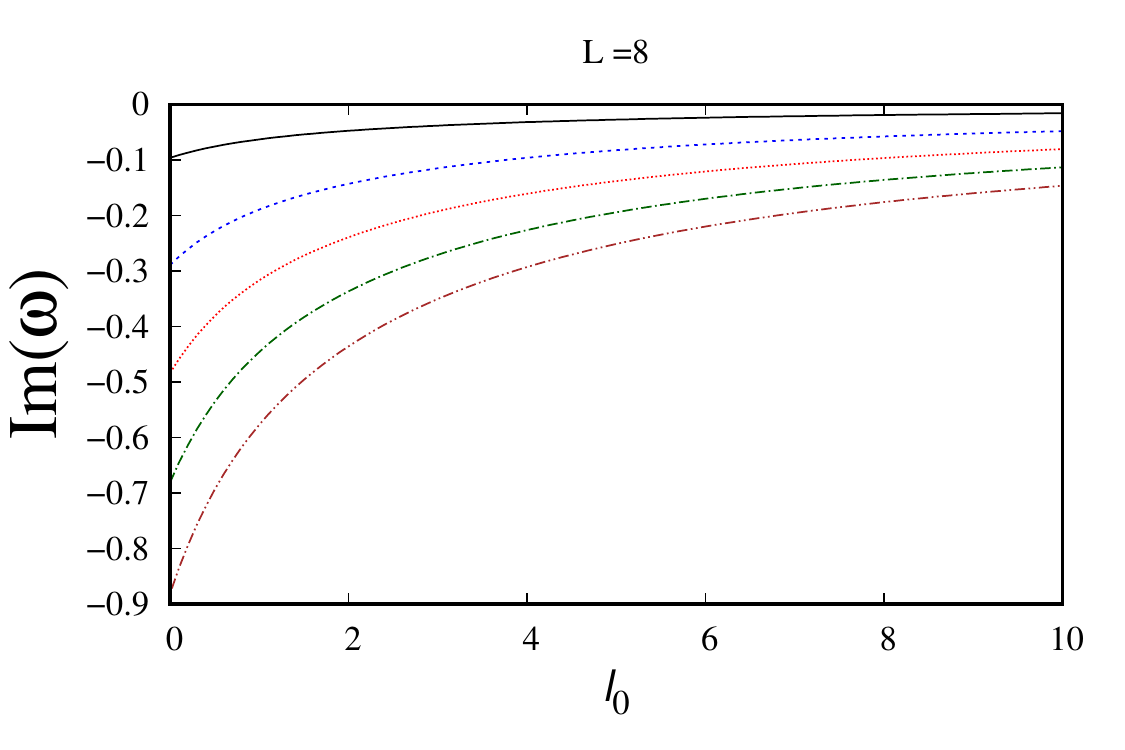}  \
\includegraphics[width=0.27\textwidth]{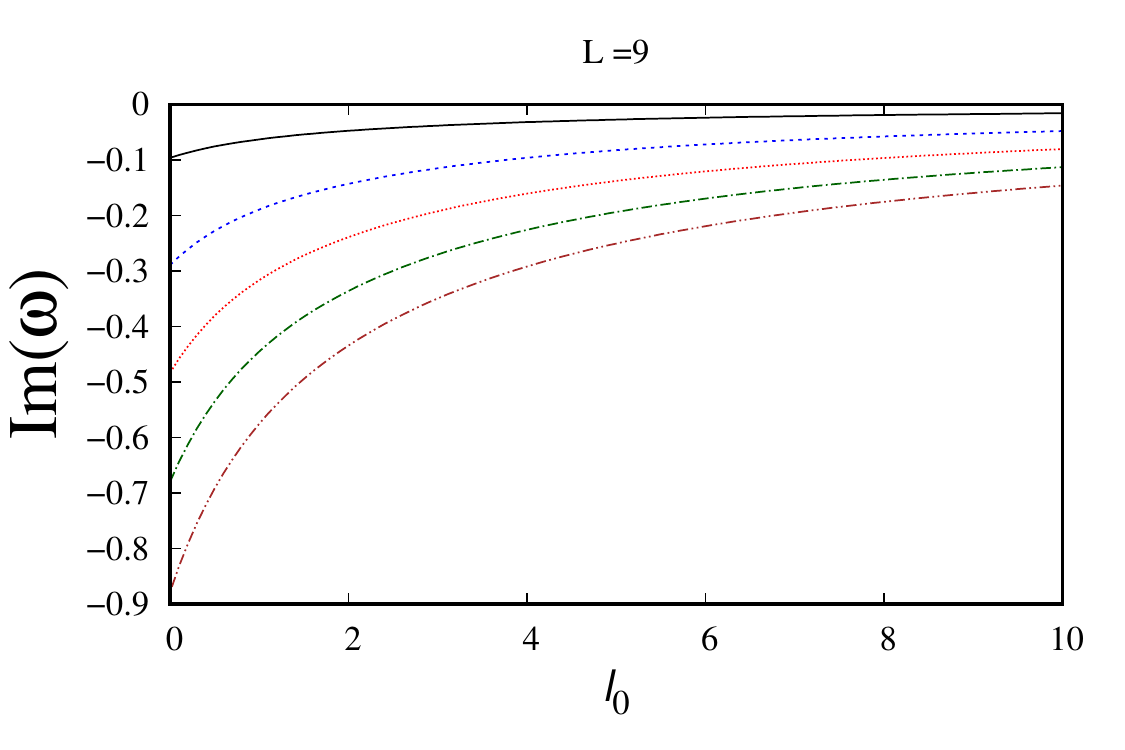}
\includegraphics[width=0.27\textwidth]{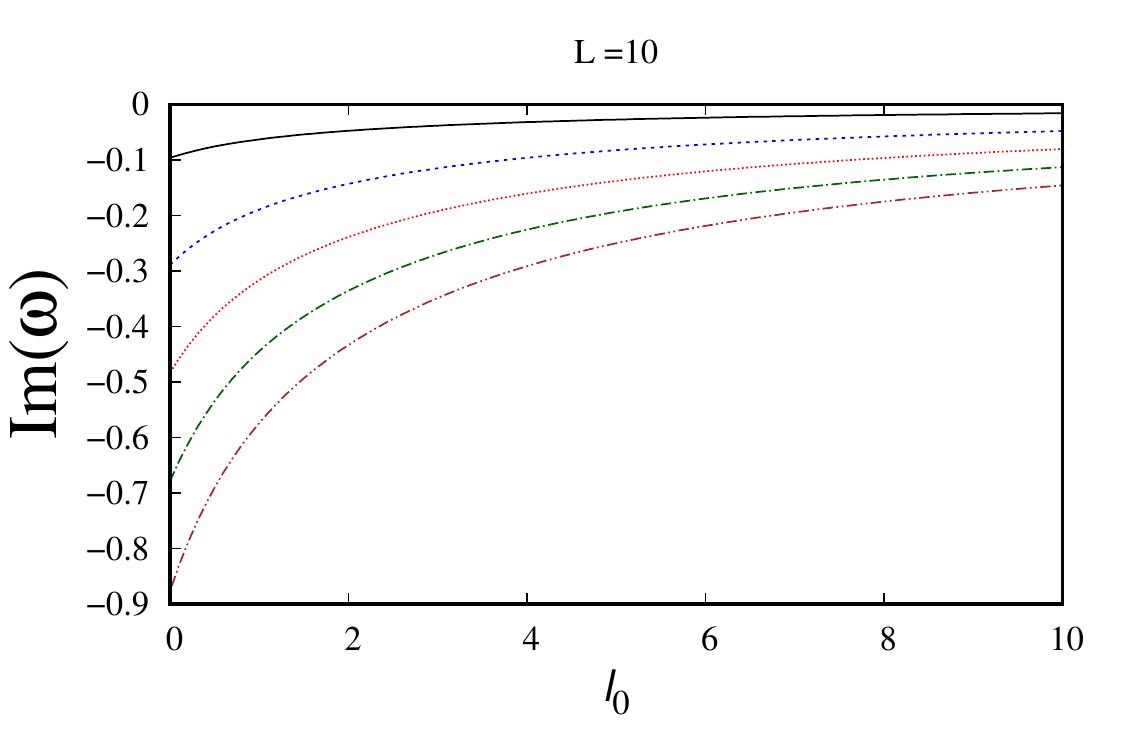}  \
\caption{\label{M4}
Imaginary part of the QNM for Model 4 as a function of the hair $\alpha$ for different values of $L$ and $n$. Each plot corresponds to a different value of $L$. The values for $n$ are $0$ (black line), $1$ (blue line), $2$ (red line), $3$ (green line), $4$ (brown line). }
\end{figure*}

\section{Conclusion}
In this work, we computed the frequencies of the quasinormal modes through the $13^{th}$ order WKB approach of four models of hairy black holes. All the results were shown as a function of the primary hair parameter of the black holes. All the plots were made by varying the values of the primary hair parameters in a step of 0.1 in $\alpha,\ell_{0}\in(-50,50)$. However, in this work we only showed the results for $\alpha,\ell_{0}\in(-10,10)$ given that such an interval contains all the information we required for the discussion.  We found that (except for some multipole parameters in Model 1) all the black holes are stable under the perturbation for the values under consideration in the sense that the imaginary part of the quasinormal mode frequencies is always negative. Besides, we obtained that for a fixed value of the multipole parameter an increase of the overtone leads to an increase of both the absolute value of the imaginary part real part of the quasinormal modes. This result shows that, at a late time, the dominant oscillatory behavior is that with the least frequency. Regarding the effect that the primary hair has on the perturbation, we found that in Models 1, 3, and 4, the damping factor diminishes as the parameter associated with the hair grows. However, for Model 2 the frequencies are almost constant so that, in this case, the primary hair has no effect on the stability of the black hole geometry.

 \subsection*{Acknowledgments}
EC is suported by Polygrant ${\rm N}^{\rm o}$  17459. The authors acknowledge Roman Konoplya for fruitful exchange of correspondence and for providing us the script for the computation of the QNM at $13^{th}$ order.
%
%\subsection*{Acknowledgments}
%

%\section*{References}
\bibliography{references.bib}
\bibliographystyle{unsrt}

\end{document}